\documentclass[aip,jcp,amsmath,amssymb,preprint]{revtex4-1}
\usepackage[utf8]{inputenc}
\usepackage{hyperref}
\usepackage{graphicx}
\usepackage{dcolumn}
\usepackage{bm}
\usepackage{xcolor}
\usepackage{xr}
\begin{document}

\title{Uncertainty Quantification for Predictions of Atomistic
  Neural Networks}

\author{Luis Itza Vazquez-Salazar}
\email{luisitza.vazquezsalazar@unibas.ch}
\affiliation{Department
  of Chemistry, University of Basel, Basel, Switzerland}
  
\author{Eric D. Boittier}
\affiliation{Department
  of Chemistry, University of Basel, Basel, Switzerland}

\author{Markus Meuwly}  \email{m.meuwly@unibas.ch} 
\affiliation{Department
  of Chemistry, University of Basel, Basel, Switzerland}
\affiliation{Department of Chemistry, Brown
  University, USA}%

\date{\today}

\begin{abstract}
The value of uncertainty quantification on predictions for trained
neural networks (NNs) on quantum chemical reference data is
quantitatively explored. For this, the architecture of the PhysNet NN
was suitably modified and the resulting model was evaluated with
different metrics to quantify its calibration, the quality of its
predictions, and whether prediction error and the predicted
uncertainty can be correlated. Training on the QM9 database and
evaluating data in the test set within and outside the distribution
indicate that error and uncertainty are not linearly related. However,
the observed variance provides insight into the quality of the data
used for training. Additionally, the influence of the chemical space
covered by the training data set was studied by using a biased
database. The results clarify that noise and redundancy complicate
property prediction for molecules even in cases for which changes -
such as double bond migration in two otherwise identical molecules -
are small. The model was also applied to a real database of
tautomerization reactions. Analysis of the distance between members in
feature space in combination with other parameters shows that
redundant information in the training dataset can lead to large
variances and small errors whereas the presence of similar but
unspecific information returns large errors but small variances. This
was, e.g., observed for nitro-containing aliphatic chains for which
predictions were difficult although the training set contained several
examples for nitro groups bound to aromatic molecules. The finding
underlines the importance of the composition of the training data and
provides chemical insight into how this affects the prediction
capabilities of a ML model. Finally, the presented method can be used
for information-based improvement of chemical databases for target
applications through active learning optimization.
\end{abstract}


\maketitle

\section{Introduction}
Undoubtedly machine learning (ML) models are becoming part of the
standard computational/theoretical chemistry toolbox. This is because
it is possible to develop highly accurate trained models in an
efficient manner. In chemistry, such ML models are used in various
branches ranging from the study of reactive
processes,\cite{MM.cr:2021,topfer2022double} sampling equilibrium
states,\cite{noe2019boltzmann}, the generation of accurate force
fields,\cite{manzhos:2020,MM.rkhs:2020,conte:2020,unke2021machine,unke2022accurate},
to the generation and exploration of chemical
space.\cite{Schwalbe-Koda2020,huang:2021} Nowadays, an extensive range
of robust and complex models can be
found.\cite{schutt2018schnet,smith2017ani1,gao2020torchani,ko2021fourth,unke2021spookynet}
The quality of these models is only limited by the quality and
quantity of the data used for
training.\cite{keith2021combining,unke2021machine} For the most part,
however, the focus was on obtaining more extensive and complex
databases as an extrapolation from applications in computer
science. Therefore, it is believed that more significant amounts of
data will beat the best algorithms.\cite{domingos2012few}\\

\noindent
On the other hand it has been found that even the best model can be
tricked by poor data
quality.\cite{sanders2017garbage,kilkenny2018data,canbek2022gaining,tweedie1994garbage}
For example, in malware detection it was found that ML-based models
can fail if the training data does not contain the event the model had
been designed for.\cite{sanders2017garbage,canbek2022gaining} Also,
data completeness and quality directly impact on the forecasting
capabilities of such statistical models. The notion of underperforming
models trained on low-quality data ("garbage in-garbage out") can be
traced back to Charles Babbage.\cite{babbage_2011} The ML community is
starting to notice the importance of data quality used for training
and the relevance to balance amount of data (``big data'') versus
quality of data. From other fields in Science it is known that using
biased and low-quality data in ML can result in catastrophic
outcomes\cite{geiger2021garbage} such as discrimination towards
minorities,\cite{c2019garbage} reduction in patient survival, and the
loss of billions of dollars.\cite{saha2014data} As a result of these
findings, the concept of "smart data"
emerged.\cite{iafrate2014journey,baldassarre2018big,triguero2019transforming}
Smart data describes a set of data which contains valid, well-defined
and meaningful information that can be
processed.\cite{baldassarre2018big} However, an important
consideration concerns the type of data that is required for
predicting a particular target property. Although quantum chemical
models are trained, for example, on total energies of a set of
molecules, it is not evident how to best select the most suitable
training set for most accurately predicting energy differences between
related compounds, such as structural isomers. For this, the
uncertainty on a prediction is very valuable additional information
because this would allow to specifically improve a given training set
for predictions that perform unsatisfactorily.\\

\noindent
Considering that data generation for training quantum ML models
implies the use of considerable amounts of computational
power\cite{von2018quantum,heinen2020machine,kaeser2020machine} which
increases the carbon footprint and makes the use of ML difficult for
researchers without sufficient resources, it is essential to optimize
the full workflow from conception to a trained model. With this in
mind, the concept of smart data is of paramount importance for
conceiving future ML models in chemistry. This necessity has been
considered in previous reviews about ML in
chemistry\cite{keith2021combining,unke2021machine}; however, it is
still poorly understood how the choice of training data influences the
prediction quality of a trained machine-learned model. One such effort
quantitatively assessed the impact of different commonly used quantum
chemical databases on predicting specific chemical
properties.\cite{vazquezsalazar2021} The results showed that the
predictions from the ML model are heavily affected by data redundancy
and noise implicit in the generation of the training dataset.\\

\noindent
Identifying missing/redundant information in chemical databases is a
challenging but necessary step to ensure the best performance of ML
models. In transfer learning from a lower level of quantum chemical
treatment (e.g. M{\o}ller-Plesset second order theory - MP2) to the
higher coupled cluster with singles, doubles and perturbative triples
(CCSD(T)) it has been found for the H-transfer barrier height in
malonaldehyde that it is rather the selection of geometries included
in TL than the number of additional points that leads to a
quantitatively correct model.\cite{MM.pt:2020} It is also likely that
depending on the chemical target quantity of interest the best
database differs from the content of a more generic chemical
database. Under such circumstances, uncertainty quantification (UQ) on
the prediction can provide valuable information on how prediction
quality depends on the underlying database used for training the
statistical model. UQ of quantum ML usually involves training several
models,\cite{janet2019quantitative,zheng2022toward} which incurs a
high computational cost, to obtain an average value and a
variance for a quantity of interest. However, as ML models become more prevalent in different
high-risk fields, new and efficient techniques for UQ have
emerged.\cite{gawlikowski2021survey,abdar2021review} Among them it is
of particular interest to use approaches for variance determination in
a single deterministic model because this is computationally cheaper
and can be used for active learning. This has been done recently for molecular discovery and inference for virtual screening and
  discussed the advantages for
  prioritization.\cite{soleimany2021evidential} In the present work, we explore
the advantage of using 'Deep Evidential
Regression'\cite{amini2020} that allows one to predict the variance for forecasting using a single deterministic model.\\

\noindent
The present article is structured as follows. First, the method
section describes modifications made to the PhysNet model for variance
prediction. Additionally, metrics used for hyperparameter optimization
are introduced. Then, results for model calibration and experiments
relating the error and variance depending on chemical composition of
the database are analyzed quantitatively and an application to
tautomerization is presented. Finally, the results are discussed in a
broader context.\\

\section{Methods}
\begin{figure}
    \centering
    \includegraphics[width=\textwidth]{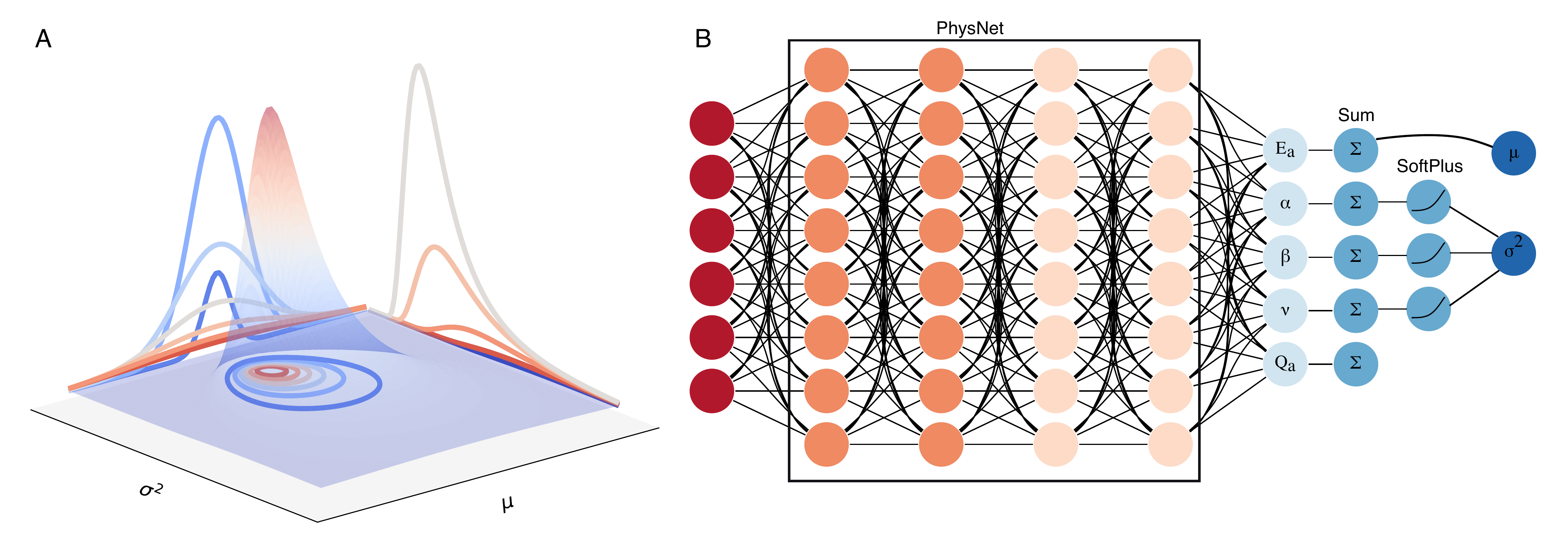}
    \caption{Modified PhysNet for uncertainty
      quantification. \textbf{A} Schematic 3D representation of the
      Negative Inverse Gamma distribution as a function of the mean
      ($\mu$) and the variance ($\sigma^{2}$) (See Equation
      \ref{eq:nig}). \textbf{B} The modified architecture of PhysNet
      for the addition of the 'evidential' layer. The input layer
      receives atomic positions, atomic numbers, charges, and
      energies. In the next step, those values are passed to the
      regular architecture of PhysNet. The final layer is modified to
      output five values ($E_{a}$, $Q_{a}$, $\alpha$, $\beta$, and
      $\nu$) per each atom in a molecule. In the next step, the values
      of the outputs are summed by each molecule. Then, the three
      extra parameters are passed to a SoftPlus activation function
      (See Equation \ref{eq:extra}). The final output of the model are
      the values for that characterize the Normal Inverse Gamma
      distribution. The mean value for the prediction (Equation
      \ref{eq:exp_mean}) corresponds to the energy of the predicted
      molecule, and the parameters to determine the variance of the
      predicted energy which can be obtained using Equations
      \ref{eq:alea_unc} and \ref{eq:epis_unc}.}
\label{fig:architecture}
\end{figure}

As a regression model, PhysNet\cite{unke2019physnet} was selected for
the present purpose. PhysNet was implemented within the PyTorch
framework \cite{paszke2019pytorch} to make it compatible with modern
GPU architectures and in line with community developments. The
original architecture of PhysNet was modified to output the energy and
three extra parameters required for the representation of the
uncertainty (Figure \ref{fig:architecture}). Following earlier
work,\cite{amini2020} it is assumed that the targets to predict (here
energies $E_i$ for samples $i$) are drawn from an independent and
identically distributed (i.i.d) Gaussian distribution with unknown
mean ($\mu$) and variance ($\sigma^{2}$) for which probabilistic
estimates are desired:
\begin{equation*}
    (E_{1},\dots,E_{N}) \approx \mathcal{N}(\mu,\sigma^{2})
\end{equation*}

\noindent
For modeling the unknown energy distribution, a prior
distribution is placed on the unknown mean ($\mu$) and variance
($\sigma^2$). Following the assumption that the values are drawn from
a Gaussian distribution, the mean can be represented by a Gaussian
distribution and the variance as an Inverse-Gamma distribution
\begin{equation*}
    \mu \sim \mathcal{N}(\gamma,\sigma^{2}\nu^{-1}), \quad \sigma^{2}\sim\Gamma^{-1}(\alpha,\beta) \\
\end{equation*}
where $\Gamma(\cdot)$ is the gamma function, $\gamma \in \mathbb{R}$,
$\nu > 0$, $\alpha>1$ and $\beta>0$.\\

\noindent
The desired posterior distribution has the form:
\begin{equation*}
    q(\mu,\sigma^{2}) = p(\mu,\sigma^{2}|E_{1},\dots,E_{N}).
\end{equation*}
where $p$ indicates a generic distribution. Following the chosen representations for mean and variance, it is
assumed that the posterior distribution can be factorized as
$q(\mu,\sigma^{2}) = q(\mu)q(\sigma^{2})$. Consequently, the joint
higher-order, evidential distribution is represented as a
Normal-Inverse Gamma distribution (Figure \ref{fig:architecture}) with
four parameters ($\mathbf{m}=\{\gamma,\nu,\alpha,\beta\}$) that
represent a distribution over the mean and the variance.
\begin{equation}
    p(\mu,\sigma^{2}|\gamma,\nu,\alpha,\beta) = \dfrac{\beta^{\alpha}
      \sqrt{\nu}}{\Gamma(\alpha)\sqrt{2\pi\sigma^{2}}} \left(
    \dfrac{1}{\sigma^{2}}\right)^{\alpha+1}
    \mathrm{Exp}\left(-\dfrac{2\beta+\nu(\gamma-\mu)^{2}}{2\sigma^{2}}\right)
\label{eq:nig}
\end{equation}
The four parameters that represent the Normal-Inverse Gamma
distribution are the output of the final layer of the trained PhysNet
model (Figure \ref{fig:architecture}) and the total predicted energy
for a molecule composed of $N$ atoms is obtained by summation of the
atomic energy contributions $E_{i}$:
\begin{equation}
    E = \sum_{i=i}^{N} E_{i}
    \label{eq:energy}
\end{equation}

\noindent
In a similar fashion, the values for the three parameters
($\nu,\alpha$, and $\beta$) that describe the distribution of the
variance for a molecule composed of $N$ atoms are obtained by
summation of the atomic contributions and are then passed to a
softplus activation function to fulfill the conditions given for the
distribution ($\gamma \in \mathbb{R}$ and $\nu$, $\alpha$, $\beta >0$)
\begin{equation}
\begin{aligned}
      \alpha = \log \left(1 + \exp \left(\sum_{i=i}^{N}
      \alpha_{i}\right) \right) + 1\\ \beta = \log\left(1 + \exp
      \left(\sum_{i=i}^{N} \beta_{i}\right) \right)\\ \nu =
      \log\left(1 + \exp \left(\sum_{i=i}^{N} \nu_{i}\right) \right)
\end{aligned}
\label{eq:extra}
\end{equation}

\noindent
Finally, the expected mean (Equation \ref{eq:exp_mean}), and the
aleatory (Equation \ref{eq:alea_unc}) and epistemic (Equation
\ref{eq:epis_unc}) uncertainty of predictions can be calculated as:
\begin{equation}
    \mathbb{E}[\mu] = \gamma
    \label{eq:exp_mean}
\end{equation}
\begin{equation}
    \mathbb{E}[\sigma^{2}]=\frac{\beta}{\alpha-1}
    \label{eq:alea_unc}
\end{equation}
\begin{equation}
    Var[\mu] = \frac{\beta}{\nu(\alpha-1)}
    \label{eq:epis_unc}
\end{equation}

\noindent
Including the new parameters in the output of the neural network
changes the loss function of the model. The new loss function consists
of a dual-objective loss $\mathcal{L}(x)$ with two terms: the first
term maximizes model fitting and the second penalizes incorrect
predictions according to
\begin{equation}
    \mathcal{L}(x) = \mathcal{L}^{\rm NLL}(x)
    +\lambda(\mathcal{L}^{\rm R}(x) - \varepsilon)
    \label{eq:loss_funct}
\end{equation}

\noindent
In equation \ref{eq:loss_funct}, the first term corresponds to the
negative log-likelihood (NLL) of the model evidence that can be
represented as a Student-$t$ distribution (Equation \ref{eq:loss_nll})
\begin{equation}
    \mathcal{L}^{\rm NLL}(x) = \frac{1}{2} \log
    \left(\frac{\pi}{\nu}\right) - \alpha \log(\Omega) + (\alpha +
    \frac{1}{2}) \log((y-\gamma)^{2}\nu + \Omega) + \log
    \left(\frac{\Gamma(\alpha)}{\Gamma(\alpha+\frac{1}{2})}\right)
    \label{eq:loss_nll}
\end{equation}
where $\Omega = 2\beta(1+\nu)$ and $y$ is the value predicted by the
neural network. For details of the derivation of this equation see 
Amini,2020\cite{amini2020}. The second term in Equation \ref{eq:loss_funct},
$\mathcal{L}^{\rm R}(x)$, corresponds to a regularizer that minimizes
the evidence for incorrect predictions (Equation \ref{eq:loss_ref}).
\begin{equation}
    \mathcal{L}^{\rm R}(x) = |y-\gamma|\cdot(2\nu + \alpha)
    \label{eq:loss_ref}
\end{equation}
The hyperparameter $\lambda$ controls the influence of uncertainty
inflation on the model fit and can be calibrated to obtain more
confident predictions. For $\lambda=0$, the model is
overconfident. i.e. results are less likely to be
correct. Alternatively, for $\lambda>0$, the variance is inflated,
resulting in underconfident predictions.\\

\noindent
The neural network architecture was that of standard PhysNet, with 5
modules consisting of 2 residual atomic modules and 3 residual
interaction modules. Finally, the result is pooled into one residual
output module. The number of radial basis functions was kept at 64,
and the dimensionality of the feature space was 128. Electrostatic and
dispersion corrections were not used for the training to keep the
model as simple as possible. All other parameters were identical to
the standard version of PhysNet\cite{unke2019physnet}, unless
mentioned otherwise.\\

\noindent
For training, a batch size of 32 and a learning rate of 0.001 were
used. An exponential learning rate scheduler with a decay factor of
0.1 every 1000 steps and the ADAM optimizer\cite{kingma2014adam} with
a weight decay of 0.1 were employed. An exponential moving average for
all the parameters was used to prevent overfitting. A validation step
was performed every five epochs.\\

\subsection{Hyperparameter Optimization}
The hyperparameter $\lambda$ in equation \ref{eq:loss_funct} was
optimized by training a range of models with different values of
$\lambda$, using a portion of the QM9 dataset consisting of 31250,
25000 structures for training, 3125 for validation and the remaining
3125 for testing. Models were trained for 1000 epochs. The values for
$ \lambda$ considered were 0.01, 0.1, 0.2, 0.4, 0.5, 0.75, 1.0, 1.5,
and 2.0.\\

\subsection{Metrics for Model Assessment and Classification}
In order to compare the performance/quality of the
trained models, suitable metrics are required. These metrics are used
to select the best value for the hyperparameter $\lambda$. Different
metrics that have been reported in the
literature\cite{levi2019evaluating,tran2020methods,busk2021calibrated}
were evaluated.\\

\noindent
The first metric considered is the Root Mean Variance (RMV) defined
as:
\begin{equation}
    \mathrm{RMV}(j) = \sqrt{\dfrac{1}{|B_{j}|}\sum_{t \in B_{j}}
      \sigma^2_{t}}
    \label{eq:rmv}
\end{equation}
Here, $\sigma_{t}^{2}$ is the variance in the $j-$th bin $B_{j}$. For the construction of the bins $B_{j}$ the data
is first ranked with respect to the variance and then split in $N$
bins $\{ B_{j}\}^{N}_{j=1}$.\\

\noindent
The next metric was the empirical Root Mean Squared Error
(RMSE):
\begin{equation}
    \mathrm{RMSE}(j) = \sqrt{\dfrac{1}{|B_{j}|}\sum_{t \in B_{j}} (y_{i}-\hat{y}_{t})^{2}}
    \label{eq:rmse}
\end{equation}
where $y_i$ is the i-th prediction and $\hat{y}_{t}$ is the average
value of the prediction in a bin $B_{j}$. Using equations \ref{eq:rmv}
and \ref{eq:rmse}, the Expected Normalized Calibration Error (ENCE):
\begin{equation}
    \mathrm{ENCE} = \dfrac{1}{M}\sum_{j=1}^{M}\dfrac{|\mathrm{RMV}(j)-\mathrm{RMSE}(j)|}{\mathrm{RMV}(j)}
    \label{eq:ence}
\end{equation}
can be obtained. Additionally, it is possible to quantify the
dispersion of the predicted uncertainties for which the Coefficient of
Variation ($C_{v}$) is
\begin{equation}
    C_{v} =
    \dfrac{1}{\mu_{\sigma}}\sqrt{\dfrac{1}{M-1}\sum_{i=1}^{M}(\sigma_{i}-\mu_{\sigma})^{2}}
    \label{eq:cv}
\end{equation}
In equation \ref{eq:cv}, $\mu_{\sigma}$ is the mean predicted standard
deviation and $\sigma_{t}$ is the predicted standard deviation for $M$ samples.\\

\noindent
The last metric used for the characterization of the predicted
variance of the tested models is the 'sharpness'
\begin{equation}
    {\rm sha} =
    \dfrac{1}{N}\sum_{i=1}^{N} var(F_{n})
    \label{eq:sha}
\end{equation}
In equation \ref{eq:sha}, the value $var(F_{n})$ corresponds to the
variance of the random variable with cumulative distribution function
$F$ at point $n$.\cite{tran2020methods} The purpose of this metric is
to measure how close the predicted values of the uncertainty are to a
single value.\cite{kuleshov2018accurate}\\

\noindent
In addition to the above metrics, calibration diagrams were
constructed with the help of the uncertainty toolbox
suite.\cite{chung2021uncertainty} Calibration diagrams report the
frequency of correctly predicted values in each interval relative to
the predicted fraction of points in that
interval.\cite{tran2020methods,pascalunc2022a} Another interpretation
of the calibration diagram is to quantify the 'honesty' of a model by
displaying the true probability in which a random variable is observed
below a given quantile; if a model is calibrated this probability
should be equal to the expected probability in that
quantile.\cite{chung2021uncertainty}\\

\noindent
The results obtained for the test dataset were then classified into
four different categories following the procedure described in Kahle
and Zipoli.\cite{kahle2022} For the present purpose, $\varepsilon^{*}
= {\rm MSE}$ (mean squared error) and $\sigma^{*}=\mathrm{MV}$ (mean
variance), and the following classes were distinguished:
\begin{itemize}
    \item True Positive (TP): $\varepsilon_{i}>\varepsilon^{*}$ and
      $\sigma_{i}>\sigma^{*}$. The NN identifies a molecule with a
      large error through a large variance. In this case, it is
      possible to add training samples with relevant chemical
      information to improve the prediction of the identified
      TP. Alternatively, additional samples from perturbed structures
      for a particular molecule could be added to the increase
      chemical diversity.
      
    \item False Positive (FP):
      $\varepsilon_{i}<\varepsilon^{*}$ and
      $\sigma_{i}>\sigma^{*}$ in which case the NN identifies a
      molecule as a high-error point but the prediction is correct. In
      this case, the model is underconfident about its prediction.
      
    \item True Negative (TN): $\varepsilon_{i}<\varepsilon^{*}$ and
      $\sigma_{i}<\sigma^{*}$. Here the model recognizes that a
      correct prediction is made with a small value for variance. For
      such molecules the model has sufficient information to predict
      them adequately by assigning a small variance. Therefore, the
      model does not require extra chemical information for an
      adequate prediction.
      
    \item False Negative (FN): $\varepsilon_{i}>\varepsilon^{*}$ and
      $\sigma_{i}<\sigma^{*}$. The model is confident about its
      prediction for this molecule but it actually performs poorly on
      it. One possible explanation for this behaviour is that
      molecules in this category are rare\cite{cheng2020addressing} in
      the training set. The model recognizes them with a small
      variance but because there is not sufficient information the
      target property (here energy) can not be predicted correctly.
\end{itemize}
In the above classifications, $i$ refers to a particular molecule
considered for the evaluation. The classification relies on the
important assumption that the MSE and the MV are comparable in
magnitude which implies that the variance predicted by the model is a
meaningful approximation to the error in the prediction. A second
desired requirement is to assure the validity of the classification
procedure and that the obtained variance is meaningful is that
$\rm{MSE}>\rm{MV}$. This requirement is a consequence of the
bias-variance decomposition of the squared
error\cite{hastie2009elements}
\begin{equation}
\begin{aligned}
  \mathbb{E}(\mathrm{MSE}){} & = \mathbb{E}[(y(x)-\mu(x))^{2}|_{x=x_{0}}]
  \\ & = \underbrace{\sigma^{2}}_{\text{Irreducible Error}} +
  \underbrace{[\mathbb{E}\mu(x_{0})-y(x_{0})]^{2}}_{\text{$Bias^2$}} +
  \underbrace{\mathbb{E}[\mu(x_{0})-\mathbb{E}\mu(x_{0})]^2}_{\text{Variance}}
\end{aligned}
\label{eq:var_bias}
\end{equation}
Equation \ref{eq:var_bias} states that the expected value
($\mathbb{E}$) of the MSE consists of three terms: the irreducible
error, the bias, and the variance. Therefore, the MSE will always be
smaller than the variance except for the case that $\mu(x)=y$ for
which those quantities are equal.\cite{schervish2014probability}.\\

\noindent
As a measure of the overall performance of the model, the {\it
  accuracy} is determined as\cite{watt2020machine}:
\begin{equation}
    \mathrm{ACC} =
    \dfrac{N_{\mathrm{TP}}+N_{\mathrm{TN}}}{N_{\mathrm{TP}}+N_{\mathrm{FN}}+N_{\mathrm{TN}}+N_{\mathrm{FP}}}
\label{eq:acc}
\end{equation}
In equation \ref{eq:acc},
$N_{\mathrm{TP}}$, $N_{\mathrm{TN}}$, $N_{\mathrm{FP}}$, and
$N_{\mathrm{FN}}$ refers to the number of true positive, true
negative, false positive, and false negative samples,
respectively. Additionally, it is possible to compute the true
positive rate ($R_{\mathrm{TP}}$) or {\it sensitivity} as:
\begin{equation}
    R_{\mathrm{TP}} = \dfrac{N_{\mathrm{TP}}}{N_{\mathrm{TP}}+N_{\mathrm{FN}}}
    \label{eq:tpr}
\end{equation}
As a complement to equation \ref{eq:tpr}, the true positive predictive
value ($P_{\mathrm{TP}}$) or {\it precision} is
\begin{equation}
    P_{\mathrm{TP}} = \dfrac{N_{\mathrm{TP}}}{N_{\mathrm{TP}}+N_{\mathrm{FP}}}
    \label{eq:PPV}
\end{equation}

\subsection{Model Performance for Tautomerization}
As a final test, the performance of the evidential model was evaluated
using a subset of the Tautobase\cite{wahl2020tautobase}, a public
database containing 1680 pairs. Previously, those molecules were
calculated at the level of theory of the QM9
database.\cite{vazquezsalazar2021,vazquezsalazarQtauto} For the
purpose of the present work, only molecules that contain less than
nine heavy atoms were included. Three neural networks with $\lambda$
values of 0.2, 0.4, and 0.75 were trained with the QM9 database.  The
QM9 database was filtered to remove molecules containing fluorine and
those that did not pass the geometry consistency check. The size of
final database size was 110 426 molecules. That number was split on 80
\% for training, 10\% for validation and 10\% for testing. The three
models were trained for 500 epochs with the same parameters as for the
hyperparameter optimization.

\section{Results}
In this section the calibration of the network is analyzed and its
performance for different choices of the hyperparameter is
assessed. Then, an artificial bias experiment is carried out and
finally, the model is applied to the tautomerization data set. Before
detailing these results, a typical learning curve for the model is
considered in Figure \ref{sifig:fig1}. As expected, the root mean
squared error obtained for the test set decreases with increasing
number of samples.\\

\subsection{Calibration of the Neural Network}
The selection of the best value for the hyperparameter $\lambda$ can
be related to the calibration of the neural network model. Ideally, a
calibrated regression model should fulfill the
condition\cite{levi2019evaluating} that
\[
\forall \sigma : \mathbb{E}_{x,y}[(\mu(x)-y)^{2}|_{\sigma(x)^{2} = \sigma^{2}}] = \sigma^{2}
\]
\noindent
where $\mathbb{E}$ is the expected value for the squared difference of
the predicted mean evaluated at $x$ minus the observed value $y$. In
other words: the squared error for a prediction can be directly
related to the variance predicted by the
model.\cite{levi2019evaluating}\\

\begin{figure}
    \centering
    \includegraphics[scale=0.5]{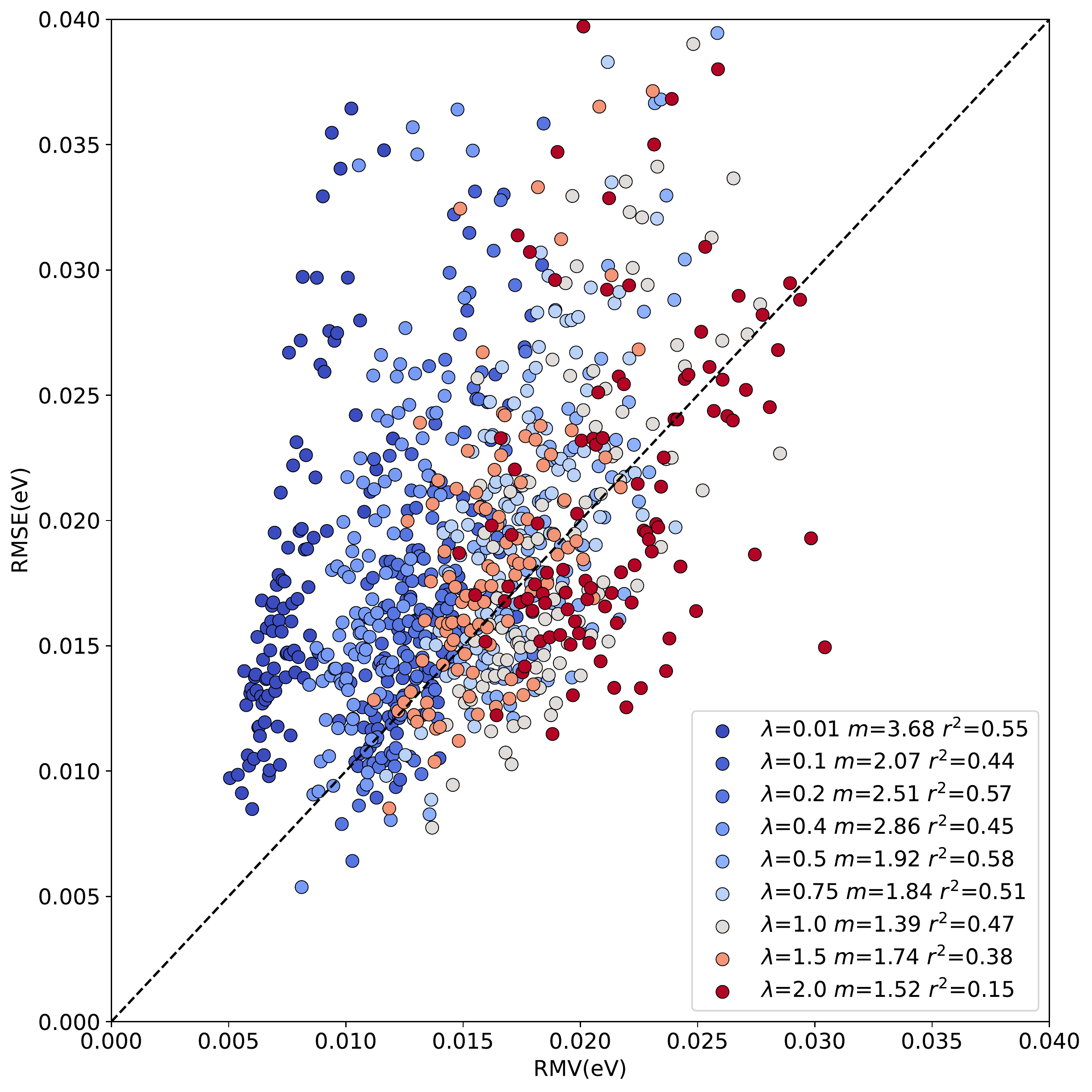}
    \caption{Empirical root mean squared error compared with the root
      mean variance of the evidential model trained on 25000
      structures from the QM9 database. The values were divided in 100
      bins ranked with respect to the predicted variance, 25 bins with
      32 samples and 75 with 31 samples were considered. The value of
      $\lambda$ together with the slope ($m$) from a linear regression
      analysis and the Pearson correlation coefficient ($r^{2}$) are
      given in the legend.}
    \label{fig:rmse_rmv}
\end{figure}

\noindent
Figure \ref{fig:rmse_rmv} compares the root mean squared error with
the root mean variance for a given number of bins ($N = 100$) and
shows that the correlation between RMSE and RMV can change between
different intervals. Additionally, the slope of the data can be used
as an indicator as to whether the model over- or underestimates the
error in the prediction. A slope closer to 1 indicates that the model
is well-calibrated. Consequently, the predicted variance can be used
as an indicator of the error with respect to the value to be
predicted. The results in Figure \ref{fig:rmse_rmv} also show that
smaller values of $\lambda=(0.01,0.2,0.4)$ result in increased slopes
of the RMSE versus RMV curve, i.e. leads to less well-calibrated
models, resulting in a model that is overconfident in its
predictions. Results that are more consistent with a slope of 1 are
obtained for $\lambda=1$. However, for all trained models it is
apparent that RMSE and RMV are not related by a ``simple'' linear
relationship as is sometimes assumed in statistical modeling.\\

\begin{figure}
    \centering
    \includegraphics[scale=0.5]{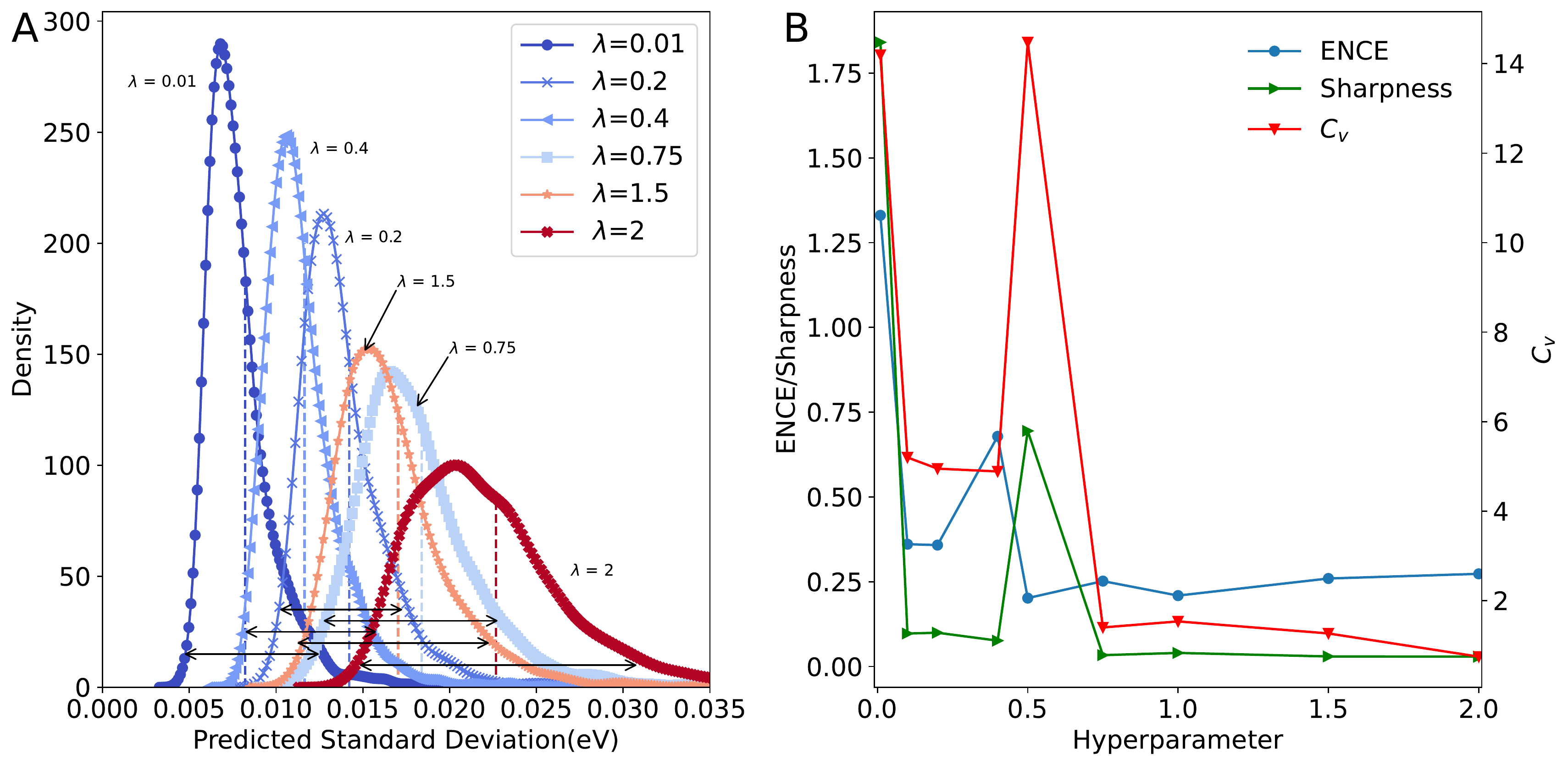}
    \caption{Metrics for the distribution of predicted
      variance. \textbf{A} Kernel density estimate of standard
      deviation($\sigma=\sqrt{\mathrm{Var}}$) for different values of
      hyperparameter $\lambda$. Values up to the 99\% percentile of
      the variance were considered. The internal arrows show the
      'width' of the distributions. Dotted lines inside the
      distribution report their sharpness. Not all distributions are
      shown for clarity. \textbf{B} Evolution of the Expected
      Normalized Calibration Error (ENCE), sharpness, and the
      Coefficient of Variation ($C_{v}$) depending on $\lambda$.}
\label{fig:dist_std}
\end{figure}

\noindent
In previous studies,\cite{tran2020methods} the dispersion of the
predicted standard deviation was considered as a measure of the
quality of a regression model. Hence a wider distribution of the
predicted standard deviation by the model is desired. To remove the
influence of pronounced outliers, Figure \ref{fig:dist_std}A shows the
distributions up to 99\% of the predicted variance. It is clear that
the center of the distribution,  and its width, depend on
$\lambda$. Larger values of the hyperparameter lead to wider
distributions. However, the displacement of the center of mass of the
distribution indicates that the standard deviation will be
consistently overestimated. Also, $p(\sigma)$ is not Gaussian but
rather resembles the inverse gamma distribution that was used as
prior for the variance.\\
name
\noindent
Predicted standard deviations from machine learned models must follow
some characteristics that help to assess the quality of model
predictions.\cite{tran2020methods} Among those characteristics, it is
expected that the distribution of the predicted variance is narrow,
i.e. will be 'sharp'. This has two objectives, the first is that the
model returns uncertainties that are as tight as possible to a
specific value.\cite{kuleshov2018accurate} With this property the
model gains confidence on its prediction. The second goal of a 'sharp'
model is that it is able to capture the
'trueness'\cite{ruscic2014uncertainty}, i.e. the distance between the
true value and the mean of the predictions, on the forecast. Another
desired characteristic is that $p(\sigma)$ is disperse and does not
return a constant value for the uncertainty which would make the model
likely to fail for predictions on molecules outside the training data
and compromise its generalizability.\\

\noindent
The previously described characteristics of the distribution of
uncertainties are related to the value of the hyperparameter $\lambda$
in the loss function (Equation \ref{eq:loss_funct}) because, as can be
seen in Figure \ref{sifig:error_var_diff}, the MSE by percentile is
independent on the choice of $\lambda$. Therefore, the model should be
calibrated by selecting a value of the hyperparameter that fulfills
the desired characteristics for the distribution of uncertainties.  \\

\noindent
Figure \ref{fig:dist_std}A shows that the spread of the distribution
of standard deviations increases with increasing $\lambda$. However,
the second desired feature for those distribution - sharpness -
decreases with increasing $\lambda$ to become almost constant for
$\lambda \geq 0.75$. In consequence of this contradictory behaviour,
it is necessary to find a value of $\lambda$ that yields an accurate
estimation of the uncertainty but it does not return a distribution of
uncertainties but rather a constant value for each case. It is
important to notice that both characteristics, sharpness and width of
the distribution, are equally important and one of them should not be
sacrificed in favour of the other.\cite{tran2020methods} In other
words: a calibrated model is characterized by uncertainty
distributions with a certain sharpness and a certain width.\\

\noindent
A deeper understanding of the difference between the error of a
predicted value and the predicted variance can be obtained through the
ENCE (Equation \ref{eq:ence}) as described in the methods
section. This metric is similar to the expected calibration error used
in classification\cite{tran2020methods}. The ENCE quantifies the probability that
the model incorrectly predicts the uncertainty of the prediction
made. Figure \ref{fig:dist_std}B reports the values of ENCE (blue
line) and shows that, typically, smaller values for ENCE are expected
for increasing hyperparameter $\lambda$. For $\lambda=0.4$, the value
of ENCE increases as opposite of the expected trend because the
predicted value of the RMSE is larger than the value for RMV for most
of the considered bins. However, it is clear that for $\lambda \geq
0.5$, the ENCE is almost constant - which indicates that, on average,
the model has a low probability to make incorrect predictions.\\

\noindent
As a complement to the ENCE metric, the coefficient of variation
($C_{\rm v}$) was also computed (red trace in Figure
\ref{fig:dist_std}B). This metric is considered to be less informative
because the dispersion of the prediction depends on the
validation/test data
distribution\cite{tran2020methods,scalia2020evaluating}. However, it
is useful to characterize the spread of standard deviations because it
is desired that the predicted uncertainties are spread and therefore
cover systems outside the training data which help to generalize the
model and make it transferable to molecules outside the training
set. Comparing the results from Figure \ref{fig:dist_std}A and the
values for $C_{\rm v}$ in Figure \ref{fig:dist_std}B, it is found that
the largest dispersion is obtained for small values of $\lambda$. This
indicates that the standard deviations for all predictions are
concentrated in a small range of values for values in the 95th
percentile of the distribution. For $\lambda \geq 0.75$ both ENCE and
$C_{\rm v}$ values do not show pronounced variation. It should be
noted that the distributions in Figure \ref{fig:dist_std}A are
restricted to the 99\% quantile of the data; on the other hand, the
values for $C_{\rm v}$ covered the whole range of data. If the
complete range of data is analyzed, it is possible to arrive at wrong
conclusions. Figure \ref{fig:dist_std}B shows that for $\lambda=0.5$,
the $C_{\rm v}$ value is large which suggests a flat distribution
(Figure \ref{sifig:all_points_dist_var}), however it should be noticed
that this behaviour arises primarily due to pronounced outliers that
impact the averages used for the calculation. However, 95\% of the
distribution is concentrated around a small range of variances as
shown in Figure \ref{fig:dist_std}A. Nevertheless, if only 95\% of the
data is studied, it is found that $\lambda \geq 0.5$ yields increased
$C_{\rm V}$ (see Figure \ref{sifig:enceandcv_95}).\\

\noindent
As shown in Figure \ref{fig:dist_std}A, the center of mass of
$P(\sigma)$ displaces to larger $\sigma$ with increasing $\lambda$.
A more detailed analysis of the difference between MSE and MV for
different percentiles of the variance was performed (Figure
\ref{sifig:error_var_diff}). Following the bias-variance decomposition
of the squared error (Equation \ref{eq:var_bias}), the bias of the
model can be quantified as a function of the different values of
$\lambda$. Figure \ref{sifig:error_var_diff} shows that the MSE is
constant regardless of the value of the hyperparameter $\lambda$ or
the percentile of the variance. On the other hand, the variance
increases as a function of $\lambda$ but it is constant regarding the
value of the percentile with the exception of $\lambda=1$. Thus, the
MV is larger than the MSE which is counter-intuitive in view of the
bias-variance decomposition of the squared error. Finally, it is clear
that the difference between MSE and MV decreases as the value of
$\lambda$ increases. This indicates that the assumed posterior
distribution does not correctly describe the data and, as a
consequence, it can not adequately capture the variance of a
prediction. In other words, a better "guess" of the posterior will
improve the predicted variance.\\

\begin{figure}
    \centering
    \includegraphics[scale=0.5]{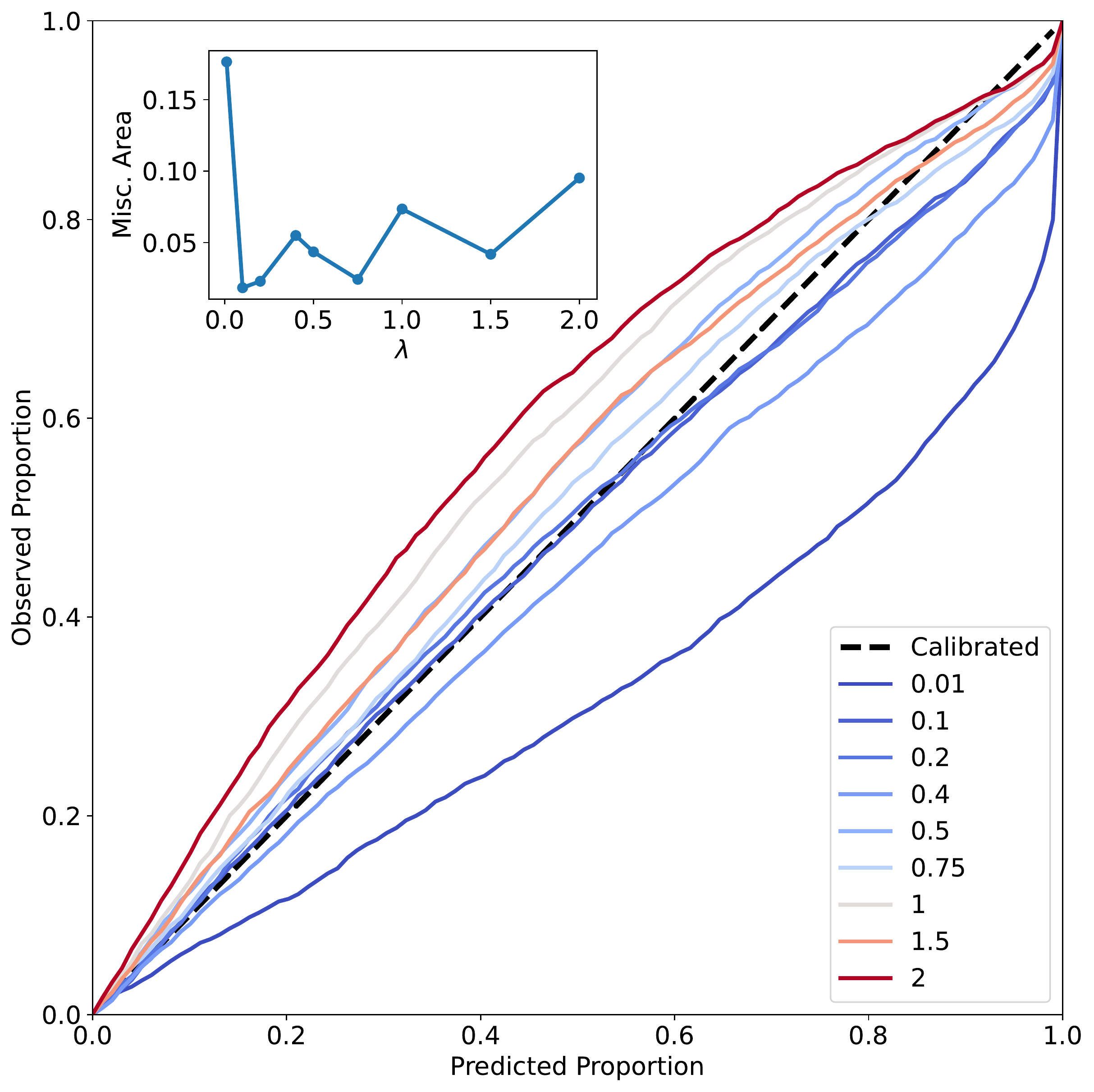}
    \caption{Calibration curves with respect to the hyperparameter
      $\lambda$. The x-axis shows the predicted probability to obtain
      the correct value for the error in a given percentile, the
      y-axis shows the true probability. The trend line shows the
      behavior of a perfectly calibrated model. Inside the plot, the
      area between the curve and the trend line, also called
      Miscalibration Area, is shown as a function of the
      hyperparameter $\lambda$. A smaller miscalibration area
      indicates a better model.}
    \label{fig:cal_curves}
\end{figure}

\noindent
A common method to judge whether a model is well-calibrated is by
considering the calibration curves described in the methods
section. The results in Figure \ref{fig:cal_curves} show that, as
$\lambda$ increases, the model is closer to the diagonal which
indicates perfect calibration. The best calibrated models are obtained
for small values of $\lambda$ ($\lambda=0.1$ and
$\lambda=0.2$). Calibration curves help to evaluate the 'honesty' of
the model predictions. Previously,\cite{soleimany2021evidential}
calibration curves were employed to select a suitable value for
$\lambda$ using the SchNet architecture\cite{schutt2018schnetpack} for
QM9. These results largely agree with what is found here with
$\lambda=0.1$ and $\lambda=0.2$ as the best values. Although
calibration curves are extensively used in the literature to assess
the quality of uncertainty predictions by ML models, they also have
weaknesses that complicate their use. For example, it was
reported\cite{levi2019evaluating} that perfect calibration is possible
for a model even if the output values are independent of the observed
error. Furthermore, it was noticed\cite{levi2019evaluating} that
calibration curves work adequately when the uncertainty prediction is
degenerate (i.e. all the output distributions have the same variance)
which is not the desired behavior. In addition to this, it was found
that the shape of these curves can be misleading because there are
percentiles for which the model under- or overestimates the
uncertainty. Then the calibration curves need to be complemented with
additional metrics for putting their interpretation in
perspective. Here, the analysis of calibration curves was complemented
by using the miscalibration area (the area between the calibration
curve and the diagonal representing perfect calibration). Using this
metric, it is clear that $\lambda$ values of 0.75 have a performance
as good as $\lambda=0.1$ and $\lambda=0.2$.\\

\subsection{Classification of Predictions}
The effect of bias in the training set for PhysNet-type models was
previously found to negatively impact prediction capabilities across
chemical space.\cite{vazquezsalazar2021} In the context of uncertainty
quantification, it is also of interest to understand how the predicted
variance can be related to the error in the prediction for an
individual prediction. For this, the relationship between the
predicted variance and the error of prediction was studied following a
classification scheme, see methods section. To this end, the subset of
QM9 used for hyperparameter optimization was considered. Then the
molecules in the test set were evaluated with the models trained with
different values of the hyperparameter $\lambda$.\\

\begin{figure}
    \centering
    \includegraphics[scale=0.55]{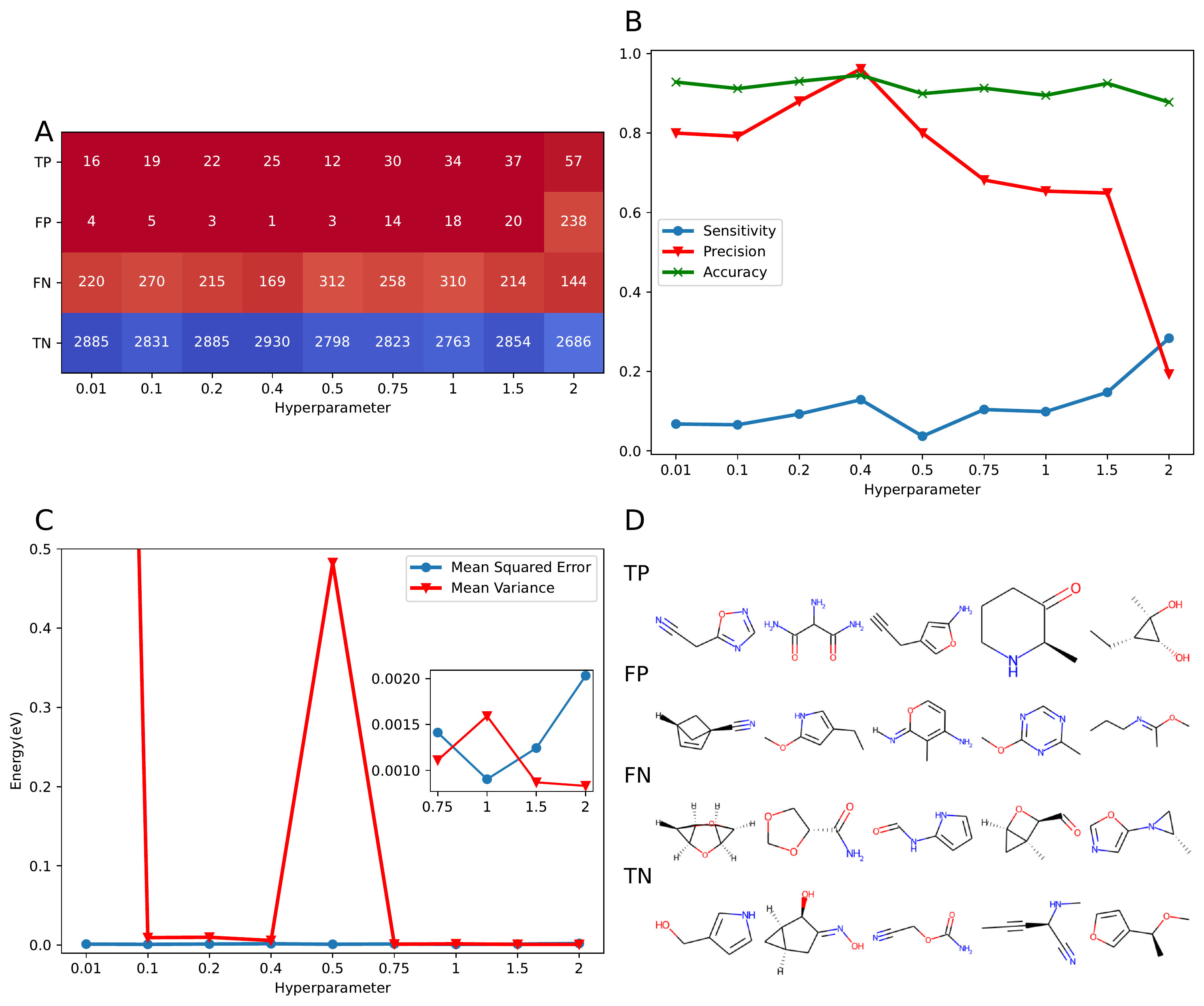}
    \caption{Results of the classification procedure. \textbf{A}
      Confusion matrix with respect to the value of the hyperparameter
      $\lambda$, inside each panel is the number of molecules that
      belong to the defined categories. The abbreviations refer to TP:
      True Positive, FP: False Positive, TN: True Negative, and FN:
      False Negative for information in how those categories are
      defined consult the methods section. \textbf{B} Accuracy (green,
      Equation \ref{eq:acc}), sensitivity (blue, Equation
      \ref{eq:tpr}), and precision (red, Equation \ref{eq:PPV})
      depending on $\lambda$. \textbf{C} The MSE and MV for the full
      set of molecules as a function of $\lambda$. The Mean Variance
      for $\lambda = 0.01$ is not shown for clarity. The inset of the
      plot shows the behavior for $\lambda \geq 0.75$. \textbf{D}
      Chemical structures of the top 5 most common molecules in each
      of the four classes.}
    \label{fig:clas_out}
\end{figure}

\noindent
For all the tested models, the largest percentage of molecules
($\approx 80\%$) was found to be True Negatives (TN), see Figure
\ref{fig:clas_out}A. This indicates that the model recognizes for most
of the samples that there is sufficient information for a correct
prediction. On the other hand, molecules classified as True Positives
(TP) correspond to samples for which predictions are difficult. Hence,
these molecules lie outside the training distribution because they are
associated with large prediction errors and the model is 'aware' of
this. As expected, the number of TP and FP increases with increasing
$\lambda$. This is a consequence of the inflation of uncertainty by
making the model less confident about its prediction which results in
misclassification of molecules because - as described before - the
error in the prediction is independent on the value of $\lambda$, see
Figure \ref{sifig:error_var_diff}. Finally, the number of False
Negative (FN) samples in the data is approximately independent on
$\lambda$. As described before, the molecules in this category contain
information on the boundary of the training distribution which
compromises the model's prediction capability. The constant number of
FN is indicative of a systematic problem that can only be corrected by
providing additional samples of similar molecules.\\

\noindent
A summary of the relationship between the four classifications in term
of model accuracy, sensitivity, and precision is given in Figure
\ref{fig:clas_out}B. In all cases the accuracy of the model is
appropriate, since the largest part ($\approx 90$\%) of the studied
samples are correctly predicted (i.e. TN) and the variance reflects
the prediction error. On the other hand, the precision of the model is
also high ($\approx 80$\%) but starts to decrease as $\lambda$
increases. In the present context, precision is a measure for the
model's capability to recognize `problematic' cases which also
correspond to a real deficiency in the model which can be assessed by
comparing the prediction with the true value and the predicted
variance. It is expected that as the model becomes more
underconfident, the precision decreases as there are more molecules
misclassified due to inflation of the uncertainty. Conversely,
sensitivity describes how many of the molecules that present a problem
in the prediction are identified by the model. Here, the sensitivity
increases for $\lambda>0.5$: as the model becomes less confident, the
probability to detect samples that are truly problematic increases. It
should, however, also be pointed out that the numerical values for
$(\epsilon^*, \sigma^*)$ to define the different categories will
impact on how the classifications impact model characteristics such as
``precision'' or ``sensitivity''.\\

\noindent
The MV and MSE for the complete set of samples as a function of
$\lambda$ are provided in Figure \ref{fig:clas_out}C. It is found that
with the exception of $\lambda=0.01$ and $\lambda=0.5$, MV and MSE are
comparable, which is a desired characteristic of the model. However,
since it is additionally desirable that MV$<$MSE the variance obtained
by the model accounts for the variance term in equation
\ref{eq:var_bias}. Therefore, the difference between MSE and MV is a
constant value that corresponds to the combination of the bias of the
model and the irreducible error. The advantage of this definition is
that the variance can be mainly attributed to the data used for
training. This provides a rational basis for further improvement of
the training data. It is noted that the condition MV$<$MSE is only
fulfilled for $\lambda = 0.75$ and $\lambda > 1.5$. A summary with the
values of all the metrics tested for calibration is given on Table
\ref{tab:resume_properties} of supporting information.\\

\noindent
Figure \ref{fig:clas_out}D and Figures \ref{sifig:TP} to
\ref{sifig:FN} present concrete molecules from each of the four
categories. Although the molecules used in the training, validation
and test sets were kept constant for the different models, the
molecules identified as outliers differed for each value of
$\lambda$. However, it is instructive to identify molecules that
appear more frequently in the various tests. These chemical structures
are studied in more detail on the following sections with the aim to
identify systematic errors and sampling problems and how they can be
corrected.\\

\subsection{Artificial bias experiment}
To provide a more chemically motivated analysis of predicted energies
and associated variances, a model was trained using the first 25k
molecules of QM9. The question addressed is whether predicted energies
and variances for molecules not used in the training of the model are
more likely to be true positives than for molecules with little
coverage in the training set. Since the structures in QM9 were derived
from graph enumeration, the order of the molecules in the database
already biases certain chemical motifs, such as rings, chains,
branched molecules and other features.\\

\begin{figure}[h!]
    \centering
    \includegraphics[height=12cm]{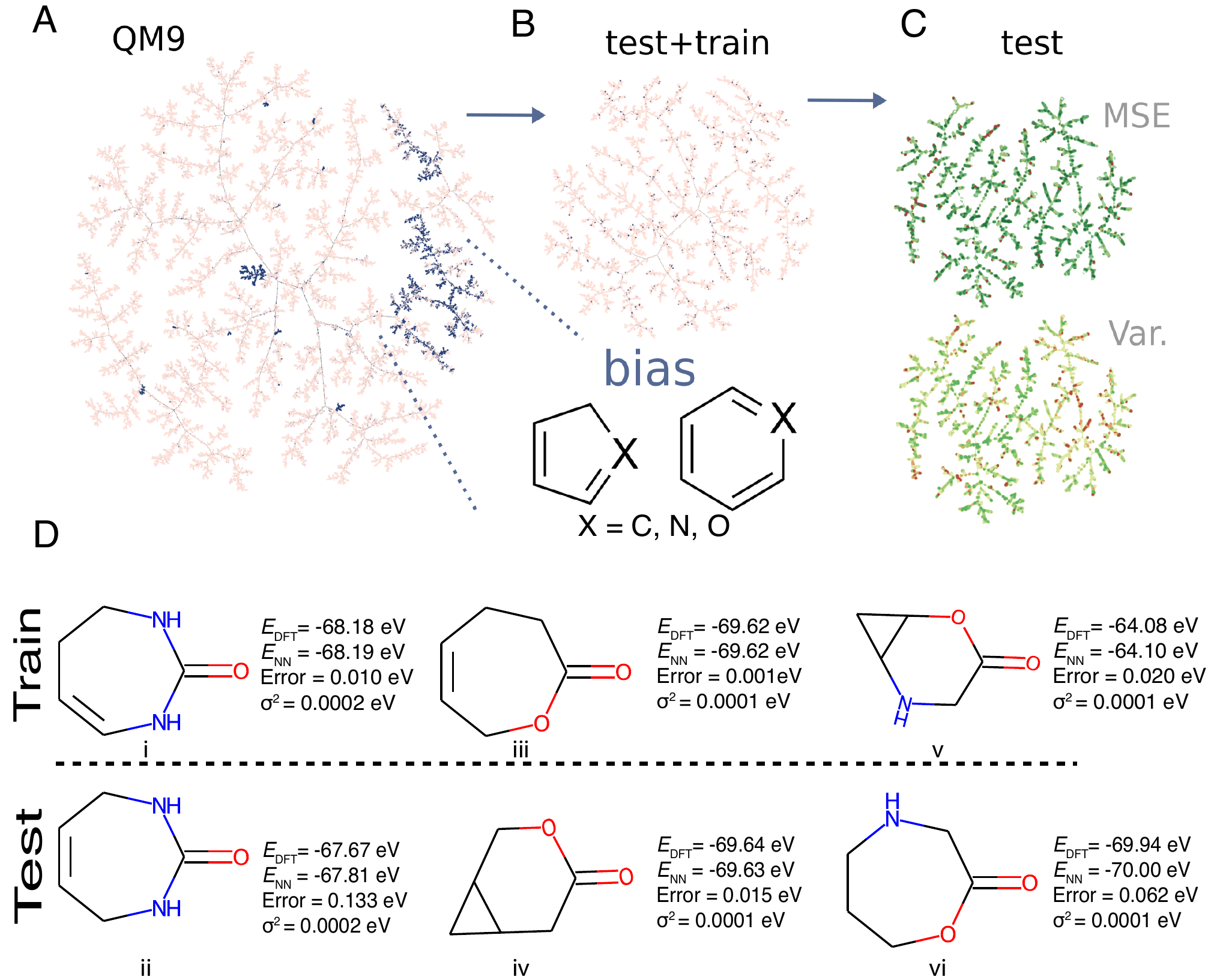}
    \caption{Artificial bias experiment. \textbf{A} TMAP of the QM9
      database. In blue the structures used for training, the inset
      shows that the selected part of the database bias the data
      towards specific chemistry, in this case, aromatic 6- and
      5-membered heterocyclic rings scaffolds. In pink, the rest of
      the structures on QM9. \textbf{B} TMAP of the reduced
      dataset. In pink the structures used for training and validation
      and in blue the selected random compounds used for
      test. \textbf{C} TMAP of the test set. On top TMAP, for the MSE
      and down the corresponding for variance. The colormaps which
      span from the minimum value (green) to $1 \, \sigma$
      (red). \textbf{D} Pairs of similar molecules (i/ii), (iii/iv),
      and (v/vi) for which one molecule was in the training set (top)
      and the related molecule was in the test set (bottom) with
      reference, prediction and difference energies displayed together
      with associated variance.}
\label{fig:chemistry}
\end{figure}

\noindent
Figure \ref{fig:chemistry}A reports the Tree MAP (TMAP)
projection\cite{probst2020visualization} of the entire QM9 database
(pink) and the first 25k molecules (blue). TMAP is a dimensionality
reduction technique with good locality-preserving properties for high
dimensional data such as molecular fingerprints.  Analysis of the
projection suggests that, as a general structural bias, the first 25k
molecules over-represent aromatic heterocyclic, 5- and 6- membered
rings, and structures with multiple substituted heteroatoms with
regards to the relative probability of other structures also present
in QM9. \\

\noindent
For training the NN, as described in the methods section, 31500
structures were randomly split (train/validation/test of 0.8/0.1/0.1)
and a model with $\lambda = 0.4$ was trained to make predictions on
the test set. A TMAP projection of the test and train compounds is
shown in Figure \ref{fig:chemistry}C. The connectivity of the
different tree branches on the TMAP provides information about the
local similarity of the molecules where dense regions of the map
correspond to clusters of high similarity. The average degree
i.e. number of edges between one molecule and its neighbors, for the
TNs in the test set - which was the majority class ($\approx 90$\% of
the test samples) - was 2.0 compared with classes FN (169 molecules),
TP (25 molecules), and FP (1 molecule) which had average degrees of
1.7, 1.3, 1.0. The lower connectivity for FP compared
with TN indicates that ``good predictions for the right reason'' are
more likely if coverage of particular structural and/or chemical
motifs is better. Furthermore, it is observed that FPs  have a low connectivity which indicates that these molecules are 'rare' in the training set. On the other hand, the different sample sizes of the
four classes need to be kept in mind when generalizing such
conclusions.\\

\noindent
The TMAP projection of the test set in Figure \ref{fig:chemistry}B
shows the chemical similarity between specific molecules seen during
training or testing. In general, molecules identified as TPs contained
common scaffolds seen during training in combination with unusual
substituents. For example, the moiety of imidazole (a five-membered
1,3-C$_3$N$_2$ ring) was a common fragment in the training set and
lies in the biased region of chemical space depicted in Figure
\ref{fig:chemistry}A. Common true positives contained this imidazole
scaffold inside uncommon fused three ring systems. When the model
makes predictions for compounds close in chemical space to molecules
of which it has seen diverse examples in the training set, the
estimates of variance appear to be more reliable.\\

\noindent
Figure \ref{fig:chemistry}D reports three examples of false positives
(i.e. molecules with high error and low predicted variance) in the
test set. The molecules in the training set are labelled as i, iii and
v, whereas those used for prediction from the test set were ii, iv and
vi. The pair (i/ii) consists of a diazepane core that goes through a
double bond migration. Although the rest of the structure is conserved
for i and ii, the error in the prediction for molecule ii (test) is
$\approx 0.1$ eV, but the predicted variance is the same for molecules
i and ii. A possible explanation is that the model recognizes that i
and ii are similar which leads to assigning a small variance to
ii. However, this contrasts with the energy difference between
molecules i and ii which is $\approx 0.5$ eV.\\

\noindent
Pair (iii/iv) involves an oxepane ring with a carbonyl (iii) which is
in the training set and the prediction is for an oxabicycloheptane
(iv). In this case the model predicts the energy with an error of
0.015 eV. Hence, for pair (iii/iv) the information that the model has
from molecule iii, in addition to the significant presence of bicycles
in the training set, makes it easier to predict the energy for
molecule iv. Finally, pair (v/vi) is opposite to (iii/iv): training on
an Oxa-azabicycloheptane for predicting an Oxazepane. The error for
this prediction is considerably higher ($\sim 0.06$ eV). This shows
that it is easier for the NN to predict bicycles than seven-membered
rings and reflects the fact that there are more bicycles in the
training set than seven-membered rings. An intriguing aspect of the
totality of molecules shown in Figure \ref{fig:chemistry}D is that
they all have the same number of heavy atoms, and that they share
multiple structural and bonding motifs. This may be the reason why the
model assigns a small variance to all of them because the NN is primed
to make best use of structural information at the training
stage. However, additional tests are required to further generalize
this.\\

\noindent
Similarly, cases where a ring was expanded or contracted by a single
atom between molecules in the training and test set commonly resulted
in similar failure modes due to over-confidence.  This observation is
particularly interesting because it suggests that the model might be
overconfident when predicting compounds it has seen sparse but highly
similar examples of during training. Uncertainty quantification, in
this conception, is effective at predicting in-distribution errors,
however, out-of-distribution errors are not as easily quantified by
this model.\\

\subsection{Tautomerization set}
As another application of how uncertainty quantification can be used,
the prediction of energy of tautomer pairs was
considered. Tautomerization is a form of reversible isomerization
involving the rearrangement of a charged leaving group within a
molecule.\cite{wilkinson1997iupac} The structures of the molecules
involved in a tautomeric pair (A/B) only differ little which makes
this an ideal application for the present developments. For the study
of tautomeric pairs, three NN models with different values of
$\lambda=0.2,0.4,0.75$ were trained with QM9 database as described on
the methods section. The test molecules considered come from the
Tautobase database.\cite{vazquezsalazarQtauto} For the purpose of this
work, only molecules with less than nine heavy atoms (C, N and O) were
tested. A total of 442 pairs (884 molecules) was evaluated.\\

\begin{figure}
    \centering
    \includegraphics[width=\textwidth]{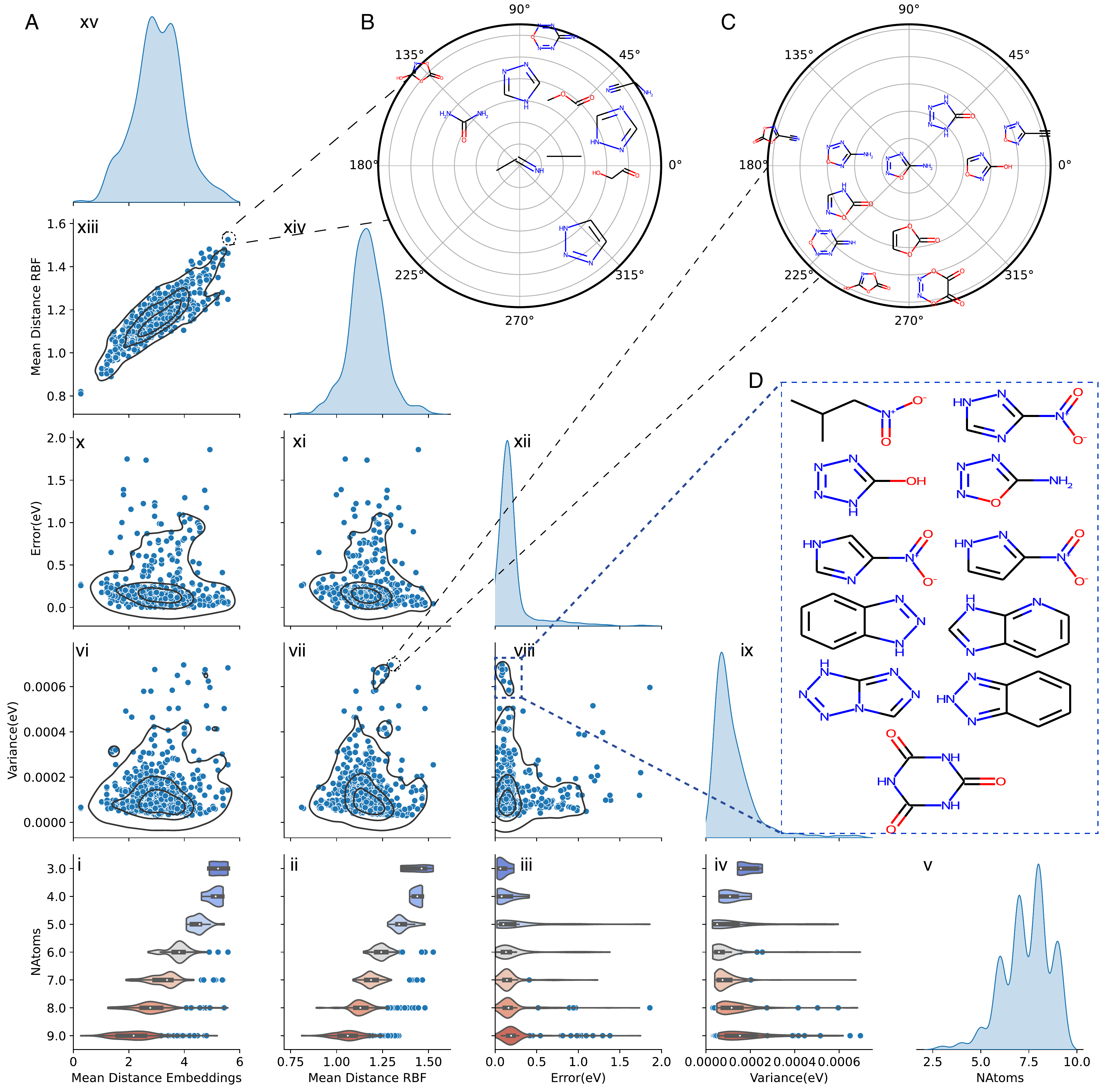}
    \caption{\textbf{A} Overview of the comparison between different
      results for the evaluation of molecules on the tautobase for
      $\lambda=0.75$ up to the 95th percentile. The diagonal of the
      figure shows the kernel density estimate of the considered
      properties (Mean Distance Embeddings, Mean Distance RBF, Error
      (in eV), Variance (in eV) and Number of Atoms). For each of the
      panels a correlation plot between the variable and a 2D kernel
      density estimate is shown. On the last row, violin plots for the
      different considered properties with respect to the number of
      atoms is shown. Similar plots for $\lambda=0.2$ and
      $\lambda=0.4$ can be found on the Supporting
      Information. \textbf{B} Radial plot of the ten closest molecules
      of the training set on feature space for the molecule in
      tautobase with the largest distance in embedding and RBF space.
      \textbf{C} Radial plot of the ten closest molecules for the
      molecule in tautobase with the largest predicted variance and
      the largest distance in RBF space.  \textbf{D} Examples of
      molecules with large predicted variance and small error.}
\label{fig:pairplot}
\end{figure}

\noindent
The training of PhysNet involves learning of the Atomic Embeddings
(AtE) and the centers and widths of the Radial Basis Functions
(RBF). These features encode the chemical environment around each atom
and therefore contain the ``chemical information'' about a
molecule. This opens the possibility to further analyze the potential
relationship contained in the learned parameters to the information
about the chemical space contained in the training dataset and how it
compares with the chemical space of the test molecules that are the
target for prediction. Hence, for the following the mean distances
between each of the tested molecules and the molecules in the training
set of the database for $\langle {\rm AtE} \rangle$ and $\langle {\rm
  RBF} \rangle$ were determined according to the procedure described
in supporting information, see SI Section
\ref{sisect:distance}. Figure \ref{fig:pairplot} shows the results for
the relationship between the mean distance of the AtE and RBF, the
error, variance and number of atoms for the molecules in the
tautobase.\\

\noindent
The bottom row of Figure \ref{fig:pairplot}A (panels i to v) report
$\langle {\rm AtE} \rangle$ and $\langle {\rm RBF} \rangle$, the
prediction errors and associated variances sorted by the number of
heavy atoms $N=3$ to 9 together with the distribution $P(N)$. The
dependence of $\langle {\rm AtE} \rangle$ and $\langle {\rm RBF}
\rangle$ on $N$ shows that with decreasing number of heavy atoms the
mean distance with respect to the molecules with the same number of
atoms increases (Figure \ref{fig:pairplot}A i and ii). Additionally,
the violin plots in Figure \ref{fig:pairplot}A i and ii show that the
mean distance values are more spread as the number of atoms
increases. One explanation for these results is that the available
chemical space to explore increases with $N$ which is also reflected
in the number of samples with a given number of heavy atoms in the
training dataset; consequently, the distance between the molecules
with a low number of atoms increases. In other words, a larger
molecule explores chemical space more extensively in terms of chemical
environments, atom types, bonding patterns and other characteristics
of chemical space. The relationship between error and the number of
atoms illustrates how the smaller mean distance in RBF and AtE leads
to a smaller error. Furthermore, the number of outliers also scales
with the size of the molecules. Comparing error and variance by the
number of heavy atoms, it is clear that for up to 5 atoms they behave
similarly (Figure \ref{fig:pairplot}A iii and iv). From Figure
\ref{fig:pairplot}A iii, it is clear that the error distribution
shifts with increasing number of atoms in the molecule. For the center
of mass of the predicted variance distribution (Figure
\ref{fig:pairplot}A iv) is at a high value and progressively decreases
until 5 heavy atoms to increase again. It should be noted that the
number of outliers for error and variance increases with the number of
heavy atoms which affects the displacement on the center of
mass. Finally, the spread of error and variance by the number of atoms
(Figure \ref{fig:pairplot}A iii and iv) presents similar shapes up to
8 heavy atoms. For molecules with 8 and 9 atoms, the variance is more
spread out whereas the error distribution is more compact.\\

\noindent
Panels vi, vii, x, and xi in Figure \ref{fig:pairplot}A show that
variance and error are similarly distributed depending on $\langle
{\rm AtE} \rangle$ and $\langle {\rm RBF} \rangle$, respectively. For
the entire range of $\langle {\rm AtE} \rangle$ and $\langle {\rm RBF}
\rangle$ low variance ($< 0.0002$ eV) and low prediction errors ($<
0.25$ eV) are found. Increased variance ($\sim 0.0005$ eV) is
associated with both, larger $\langle {\rm AtE} \rangle$ and $\langle
{\rm RBF} \rangle$ whereas larger prediction errors ($> 1.0$ eV) are
found for intermediate to large $1.0 \leq \langle {\rm RBF} \rangle
\leq 1.5$. This similarity is also reflected in a near-linear
relationship between $\langle {\rm AtE} \rangle$ and $\langle {\rm RBF}
\rangle$ reported in panel xiii of Figure \ref{fig:pairplot}A.\\

\noindent
Prediction error and variance are less well correlated for the
evaluated molecules from tautobase, see panel viii of Figure
\ref{fig:pairplot}A. This can already be anticipated when comparing
panels i and ii. With increasing $N$, the position of the maximum
error shifts monotonously to larger values whereas the variance is
higher for $N=3$, decreases until $N=6$, after which it increases
again. Hence, for tautobase and QM9 as the reference data, base error
and variance are not necessarily correlated.\\

\noindent
To gain a better understanding of the prediction performance of QM9
for molecules in the Tautobase from the point of view of feature
space, polar plots considering extreme cases were constructed, see
supporting information for technical details. Figure
\ref{fig:pairplot}B shows the case for the molecule (center) with the
largest average distance in RBF and AtE for molecules with the same
number of atoms used for training for this representation; only the
ten closest neighbours are shown. Although the molecule is relatively
simple, no structure in the training set contains sufficient and
appropriate information for a correct prediction. Despite abundant
information about similar chemical environments but with different
spatial arrangements, combination with different functional groups or
different bonding arrangements, potentially conflicting information in
the training set leads to uncertainties in the prediction. A second
example, that of the molecule with largest variance and largest
distance in RBF, is shown in Figure \ref{fig:pairplot}C. As for
molecule ii in Figure \ref{fig:chemistry}D this case also highlights
how seemingly small changes in bonding pattern, functional groups and
atom arrangements can lead to large errors. However, in this case the
abundant and similar structural information in the training set leads
to a large predicted variance. In other words, ``redundancy'' in the
training set can lead to vulnerabilities in the trained model as was
previously found for predictions based on training with the ANI-1
database compared with the much smaller ANI-1E set: despite its larger
size, predictions based on ANI-1 were less accurate than those based
on ANI-1E.\cite{vazquezsalazar2021}\\

\noindent
As a final example of the relationship between error and variance, the
chemical structures for a set of molecules with low error but high
variance is highlighted in Figure \ref{fig:pairplot}D and shows that
heterocyclic rings and bicycles are well covered in the training
set. An interesting aspect is that molecules with a nitro-group
(--NO$_2$) appear with high variance and low error. This effect can be
rationalized by considering the design of the GDB-17
Database\cite{ruddigkeit2012enumeration} which is the source of the
QM9 set: for GDB-17 aliphatic nitro groups were excluded, but aromatic
nitro groups were retained. Therefore, the trained model will have
similar information based on structural considerations but the quality
of the data in view of a molecules' energetics is low which leads to
significant variance.\\

\begin{figure}
    \centering
    \includegraphics[width=1.\textwidth]{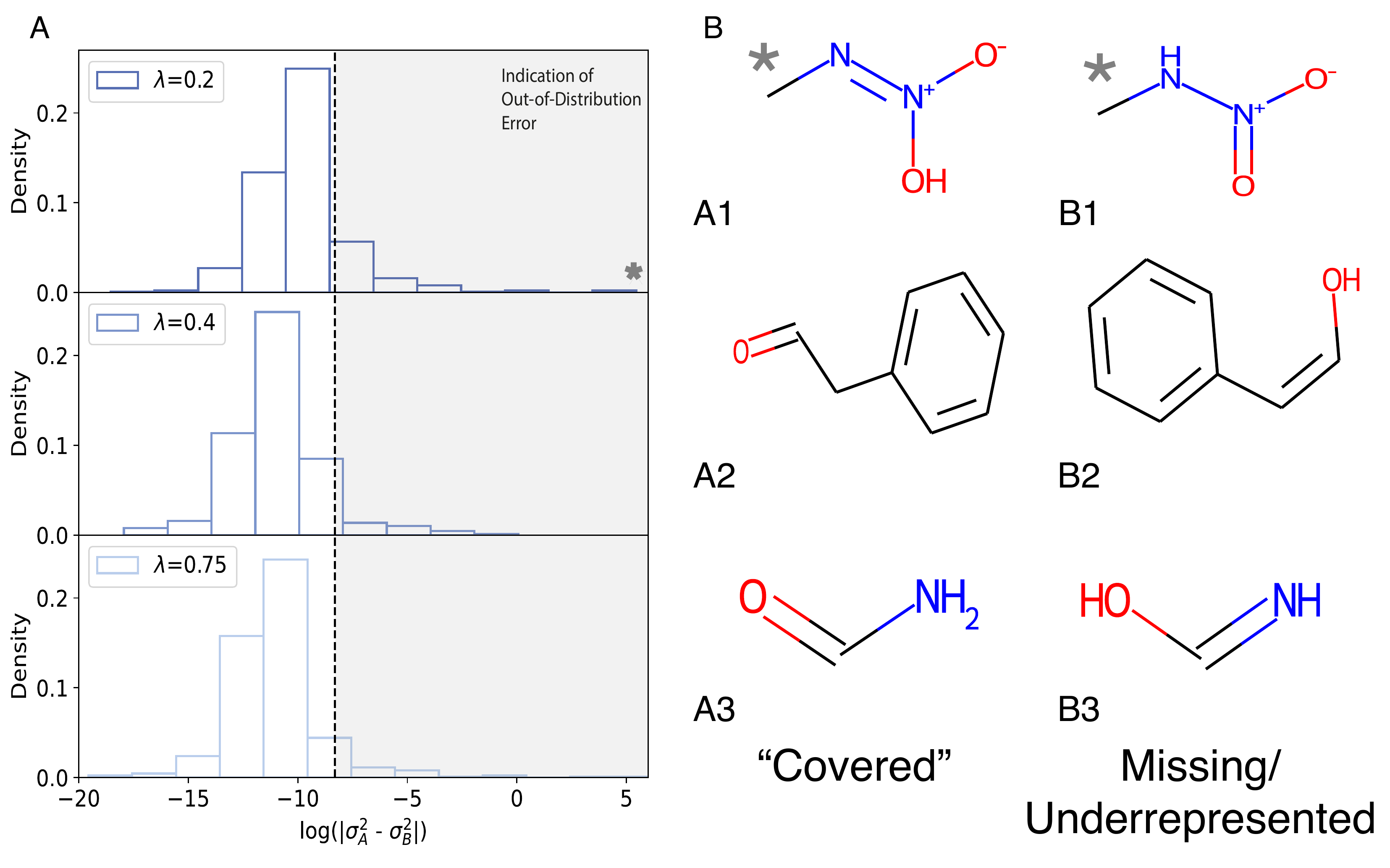}
    \caption{\textbf{A} The log distribution of differences in
      predicted variance between tautomer pairs, A (low variance) and
      B (high variance). \textbf{B} Tautomer pairs (A/B) containing
      chemical groups, nitro and vinyl alcohols, outside the training
      set (B1-3) are easily identified. The imine group in B2 was
      present in only one molecule in the training set. Numerical
      values for energies and variances are summarized in Table
      \ref{tab:tab1}.}
\label{fig:tautomers}
\end{figure}

\noindent
Finally, it is of interest to analyze tautomer pairs (A/B) for which
the difference in the predicted variance is particularly large. Figure
\ref{fig:tautomers}A reports the distribution $p(\sigma^{2}_{A} -
\sigma^{2}_{B})$ for trained models with different values of the
hyperparameter $\lambda$. First, it is found that the distribution of
variance differences depends on the value of $\lambda$. Therefore,
particularly prominent outliers can be avoided by careful evaluation
of the predictions. Secondly, large differences (star in Figure
\ref{fig:tautomers}A) in the variances can occur and indicate that the
trained models are particularly uncertain in their prediction. To
illustrate this, three tautomer pairs were identified and are analyzed
in more detail in the following. For molecules B1 to B3 it is found
that their functional groups are not present or are poorly represented
in QM9. These include the N$=$O nitro group in an aliphatic chain
(B3), vinyl alcohol (B1), and hydroxyl imine (B2, only one
representative in QM9). Furthermore, the pair (A3/B3) is
zwitterionic.\\

\begin{table}[h]
\centering
\caption{Reference energy ($E_{\rm DFT}$), predicted energy ($E_{\rm NN}$) and variance
  ($\sigma^{2}$) for selected molecules in Figure
  \ref{fig:tautomers}. All values are in eV.}
\begin{tabular}{c|c|ccc|ccc}
Molecule & $E_{\rm DFT}$     & \multicolumn{3}{c|}{$E_{\rm NN}$}       & \multicolumn{3}{c}{$\sigma^2$}  \\
\hline
$\lambda$   &          & 0.2000   & 0.4000   & 0.7500   & 0.2000   & 0.4000 & 0.7500 \\
\hline
A1       & -79.8900 & -79.6800 & -79.6900 & -79.6800 & \textbf{0.0018}   & 0.0002 & 0.0002 \\
$\Delta$    &          & \textbf{0.2100}   & 0.2000   & \textbf{0.2100}   &          &        &  \\
B1       & -79.5900 & -79.2800 & -79.4400 & -79.3600 & \textbf{0.0249}   & 0.0002 & 0.0002 \\
$\Delta$    &          & \textbf{0.3100}   & 0.1500   & 0.2300   &          &        &        \\
\hline
A2       & -23.5900 & -23.5200 & -23.5200 & -23.5200 & \textbf{0.0016}   & 0.0012 & 0.0002 \\
$\Delta$    &          &\textbf{ 0.0700}   & \textbf{0.0700}   & \textbf{0.0700}   &          &        &        \\
B2       & -23.0200 & -22.7500 & -22.8800 & -22.9100 & \textbf{0.0019}   & 0.0011 & 0.0007 \\
$\Delta$    &          & \textbf{0.2700}   & 0.1400   & 0.1100   &          &        &  \\
\hline
A3       & -30.8600 & -32.6700 & -32.5800 & -32.2000 & 0.0046   & 0.0004 & \textbf{0.2243} \\
$\Delta$     &          & \textbf{1.8100}   & 1.7200   & 1.3400   &          &        &        \\
B3       & -31.6300 & -32.0200 & -32.0800 & -32.6900 & \textbf{229.6200} & 0.0035 & 0.0033 \\
$\Delta$    &          & 0.3900   & 0.4500   & \textbf{1.0600}   &          &        &        
\end{tabular}
\label{tab:tab1}
\end{table}

\noindent
As shown in Figure \ref{fig:tautomers}B the chemical motifs and
functional groups in A1 to A3 are covered by QM9 whereas those in
their tautomeric twins (B1 to B3) are not.  For molecule B1 (vinyl
alcohol) examples are entirely absent in QM9 and the presence of
hydroxyl groups bound to sp$^2$(aromatic) carbons is not sufficient
for a reliable prediction for B1. It is also noted that the difference
$\Delta$ between the target energy ($E_{\rm DFT}$) and the predictions
($E_{\rm NN}$) are largely independent on $\lambda$ for A1 but differ
by a factor of two for B1. This is also observed for the pair (A2/B2)
for which the uncertainties are more comparable than for (A1/B1).\\

\noindent
Finally, the pair (A3/B3) poses additional challenges. First, the
variance for one value of $\lambda$ for B3 is very large and for A3
one of the variances is also unusually large, given that similar
examples to A3 are part of the training set. Secondly, although A3 is
better represented in the training set, the difference between target
value and prediction is larger than 1 eV for all models. These
observations are explained by the fact that (A3/B3) are both
zwitterionic and the uncertainty associated with B3 may in part be
related to the fact that QM9 only contains few examples of sp$^2$ NO
bonds except for a small number of hetrocylic rings which are
chemically dissimilar compounds compared with B3. Furthermore, for B3
some of the atom-atom separations (``bond lengths'') are poorly
covered by QM9. For the N--N distance, the QM9 database contains the
range from 1.2 \AA\/ to 1.4 \AA\/ (see Figure \ref{sifig:bond_dist})
whereas N--N in B3 is 1.383 \AA\/ which is a low probability region
for $p(r_{\rm NN})$. This is also the case for compound A3 although
$p(r_{\rm NN})$ has a local maximum at the corresponding N--N
separation. In conclusion, the majority of prediction problems in
Figure \ref{fig:tautomers}B can be related to origins in the
underlying chemistry. Interestingly, even a careful analysis of the
performance of a trained model on the training set (see compound A3)
may provide insight into coverage and potential limitations when
making predictions from the trained model.\\

\section{Conclusions}
The present work introduces uncertainty quantification for the
prediction of total energies and variances for molecules based on a
trained neural network. The approach is generic and it is expected
that it can be transferred to other NN-architectures and
observables.\\

\noindent
Essential findings of the present work concern the notion that single
metrics are not particularly meaningful to judge the calibration of a
trained model. Exploration and development of meaningful metrics will
benefit evidence-based inference. Also, it is not always true that
error and variance are directly related which is counter typical
expectations in statistical learning. For such findings uncertainty
quantification is essential and reveals that the nature and coverage
of the training set used for model construction plays an important
role. It is also demonstrated that mean variance and mean squared
error can behave in counter-intuitive ways which points towards
deficiencies in the assumed posterior distribution. Finally, it is
demonstrated for the case of tautomerization energies that
classification of predictions can be used to isolate problematic cases
at the prediction stage.\\

\noindent
These shortcomings can be linked to missing or poor coverage in the
training data set. Considering the mean distance in feature space, it
was found that information needs to be added in a rational fashion
because the presence of too much redundant information destabilizes
the model. This can be seen, e.g., in that large variances are
assigned to molecules that are otherwise correctly predicted. Similar
information in low quantity returns low uncertainties but high errors,
whereas similar information in large quantities results in small
errors but high predicted uncertainties. A notable example of this is
the nitro group in the training database, which is not present for
aliphatic chains but for aromatic rings. Thus, for a balanced ML-based
model for chemical exploration an equilibrium between the quantity and
the quality of the data in the database is required. This information
can be used in the future to build targeted and evidence-based
datasets for a broad range of chemical observables based on active
learning strategies and for constructing robust high-dimensional
potential energy surfaces of molecules.\\

\section*{Acknowledgments}
The authors acknowledge financial support from the Swiss National
Science Foundation (NCCR-MUST and Grant No. 200021-7117810, to MM) and
the University of Basel. LIVS acknowledges fruitful discussions with
Dr. Kai T\"opfer and Dr. Oliver Unke.

\bibliography{aipsamp}

\clearpage

\renewcommand{\thetable}{S\arabic{table}}
\renewcommand{\thefigure}{S\arabic{figure}}
\renewcommand{\thesection}{S\arabic{section}}
\renewcommand{\d}{\text{d}}
\setcounter{figure}{0}  
\setcounter{section}{0}  

\clearpage

{\bf Supporting Information: Uncertainty Quantification for
  Predictions of Atomistic Neural Networks}

\section{Calculation of the mean distance between training and test molecules in feature space.}
\label{sisect:distance}
The mean distance between molecules from the tautobase and the
molecules in QM9 was calculated as follows. First, for each of the
molecules in the test set, molecules with the same number of atoms in
the tautobase were filtered. This is necessary because the size of the
matrices for the Radial Base Functions (RBFs) and the Atomic
Embeddings (AtE) need to be of the same size. In the second step, the
pairwise distance between the test molecule and the filtered molecules
was calculated using the Euclidean norm between points as
\begin{equation}
    ||X||_{i} = \dfrac{1}{n} \sqrt{ \sum_{j=1}^{n}\sum_{k=1}^{n}(x_{j}-y^{i}_{k})^{2}}
    \label{sieq:mean_distance}
\end{equation}
where $x_{j}$ is the element in the RBF/AtE matrix of the test
molecule, $y^{i}_{k}$ is the element in the RBF/AtE matrix for each of
the retained molecules, and $n$ is the element on the RBF
matrix/embedding matrix. Therefore the value obtained from equation
\ref{sieq:mean_distance} is the mean distance between the RBF/AtE
matrix of the test molecule and the $i$th-molecule on the training
dataset.\\

\noindent
In the next step, the distance between each of the molecules and the
test molecule is averaged to obtain the average distance in feature
space (RBF and AtE) between the target molecule and the molecules with
the same number of atoms in embedding space as:
\begin{equation}
    \langle {\rm RBF/AtE} \rangle =  \dfrac{1}{N}\sum_{i=1}^{N} ||X||_{i}
\end{equation}

\subsection{Construction of polar plots}
The polar plots in Figures \ref{fig:pairplot}B, C, and
\ref{sifig:polar_maxdist} to \ref{sifig:polar_maxvar} represent a
projection of the distance between a test molecule (at the center) and
the close molecules in embedding space. The value for $||{\rm AtE}||$
and $||{\rm RBF}||$ between the test molecule and the molecule in the
training dataset obtained from \ref{sieq:mean_distance} are
transformed to polar values according to the following expressions:
\begin{equation}
    r = \sqrt{||AtE||^{2} + ||RBF||^{2}}
\end{equation}

\begin{equation}
    \theta = \arctan{\dfrac{||AtE||}{||RBF||}}
\end{equation}

\section{Tables}

\begin{table}[h]
    \centering
\begin{tabular}{c|c|c|c|c|c|c|c|c|c|c|c}
$\lambda$ & Calibration & MSE & MV & Sharpness & Dispersion & ENCE   & $C_{v}$     & MA    & Sensitivity & Precision & Accuracy \\ \hline
0.01   &   Poor         & 0.0012 & 3.3895 & 0.0078    & 0.5500     & 0.1930 & 1.3041 & 0.1765 & 0.0678      & 0.8000    & 0.9283   \\
0.1    &      Poor   & 0.0009 & 0.0094 & 0.0133    & 0.4400     & 0.1465 & 0.3196 & 0.0184 & 0.0657     & 0.7917    & 0.9120   \\
0.2    &      Regular      & 0.0013 & 0.0099 & 0.0137    & 0.5700     & 0.1416 & 0.3080 & 0.0229 & 0.0928      & 0.8800    & 0.9302   \\
0.4    &       Good    & 0.0017 & 0.0058 & 0.0113    & 0.4500     & 0.1422 & 0.6542 & 0.0549 & 0.1289      & 0.9615    & 0.9456   \\
0.5    &       Poor     & 0.0010 & 0.4827 & 0.0189    & 0.5800     & 0.1383 & 0.1775 & 0.0435 & 0.0370      & 0.8000    & 0.8992   \\
0.75   &        Good    & 0.0014 & 0.0011 & 0.0179    & 0.5100     & 0.1545 & 0.2554 & 0.0242 & 0.1042      & 0.6818    & 0.9130   \\
1      &        Poor    & 0.0009 & 0.0016 & 0.0196    & 0.4700     & 0.1649 & 0.2020 & 0.0735 & 0.0988      & 0.6538    & 0.8950   \\
1.5    &       Regular     & 0.0012 & 0.0009 & 0.0165    & 0.3800     & 0.1551 & 0.2357 & 0.0418 & 0.1474      & 0.6491    & 0.9251   \\
2      &       Poor     & 0.0020 & 0.0008 & 0.0220    & 0.1500     & 0.1684 & 0.2542 & 0.0952 & 0.2836      & 0.1932    & 0.8778  
\end{tabular}
\caption{Summary of the properties tested for calibration of the
  evaluated models. Units when necessary are in eV. MSE: Mean Square
  Error, MV: Mean Variance, ENCE: Expected Normalized Calibration
  Error, MA: Miscalibration Area. }
\label{tab:resume_properties}
\end{table}

\begin{figure}[h!]
    \centering
    \includegraphics[width=1\textwidth]{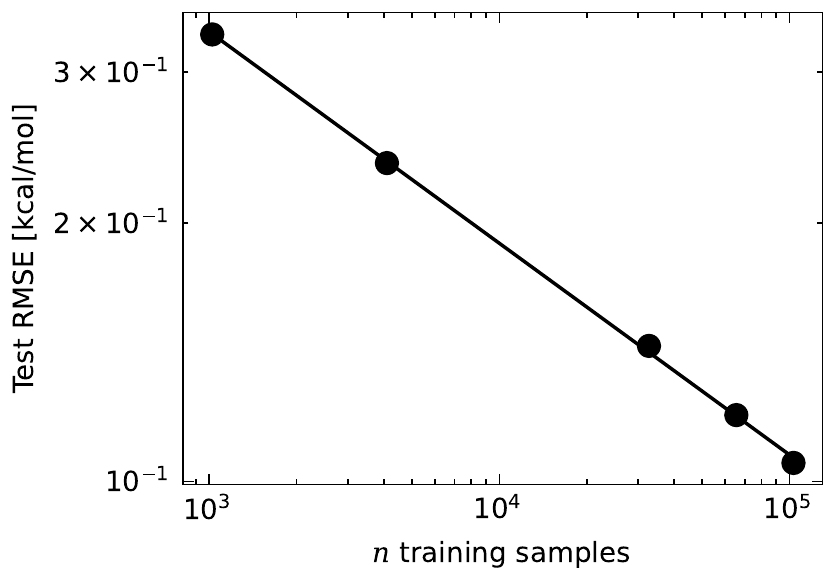}
    \caption{Learning curves showing the improvement of model with
      respect to training set size.}
\label{sifig:fig1}
\end{figure}

\section{figures}
\begin{figure}
    \centering
    \includegraphics[scale=0.7]{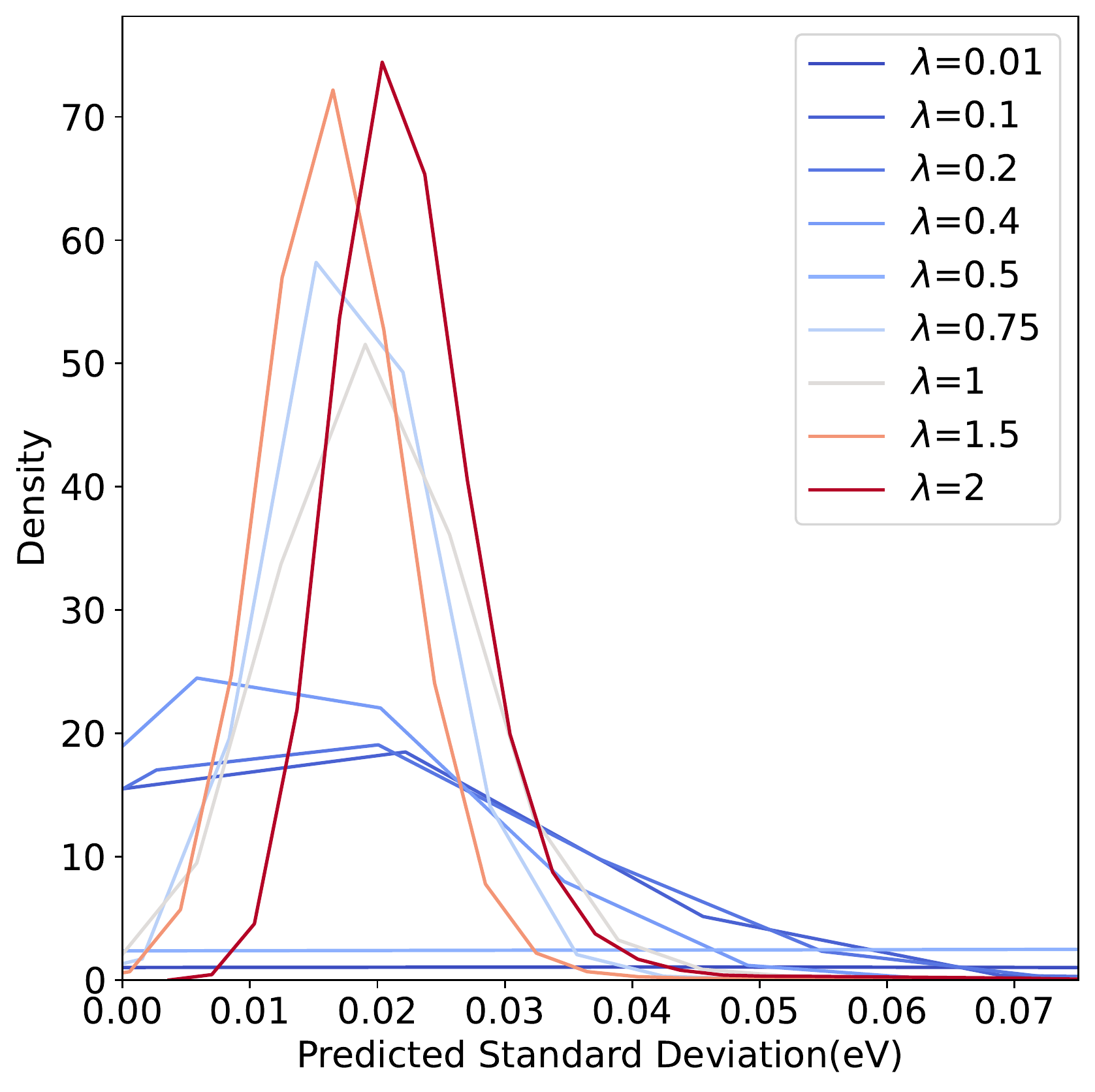}
    \caption{Distribution of standard deviation($\sigma=\sqrt{Var}$)
      for different values of hyperparameter $\lambda$.}
\label{sifig:all_points_dist_var}
\end{figure}

\begin{figure}
    \centering \includegraphics[scale=0.7]{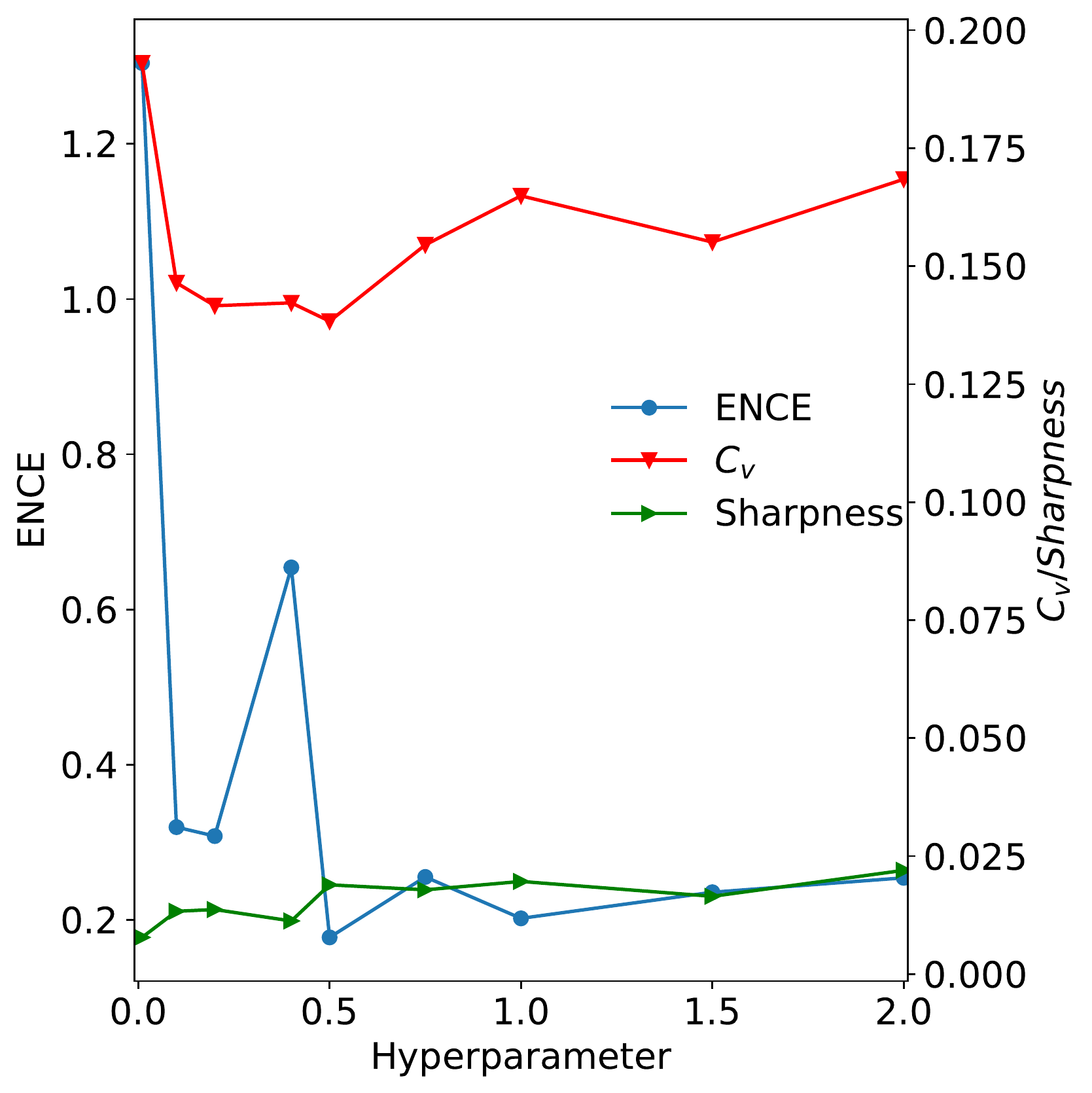}
    \caption{ Evolution of the Expected Normalized Calibration Error
      (ENCE), sharpness, and the Coefficient of Variation ($C_{v}$)
      depending on $\lambda$ for 95\% of the variance distribution.}
\label{sifig:enceandcv_95}
\end{figure}

\begin{figure}
    \centering \includegraphics[scale=0.5]{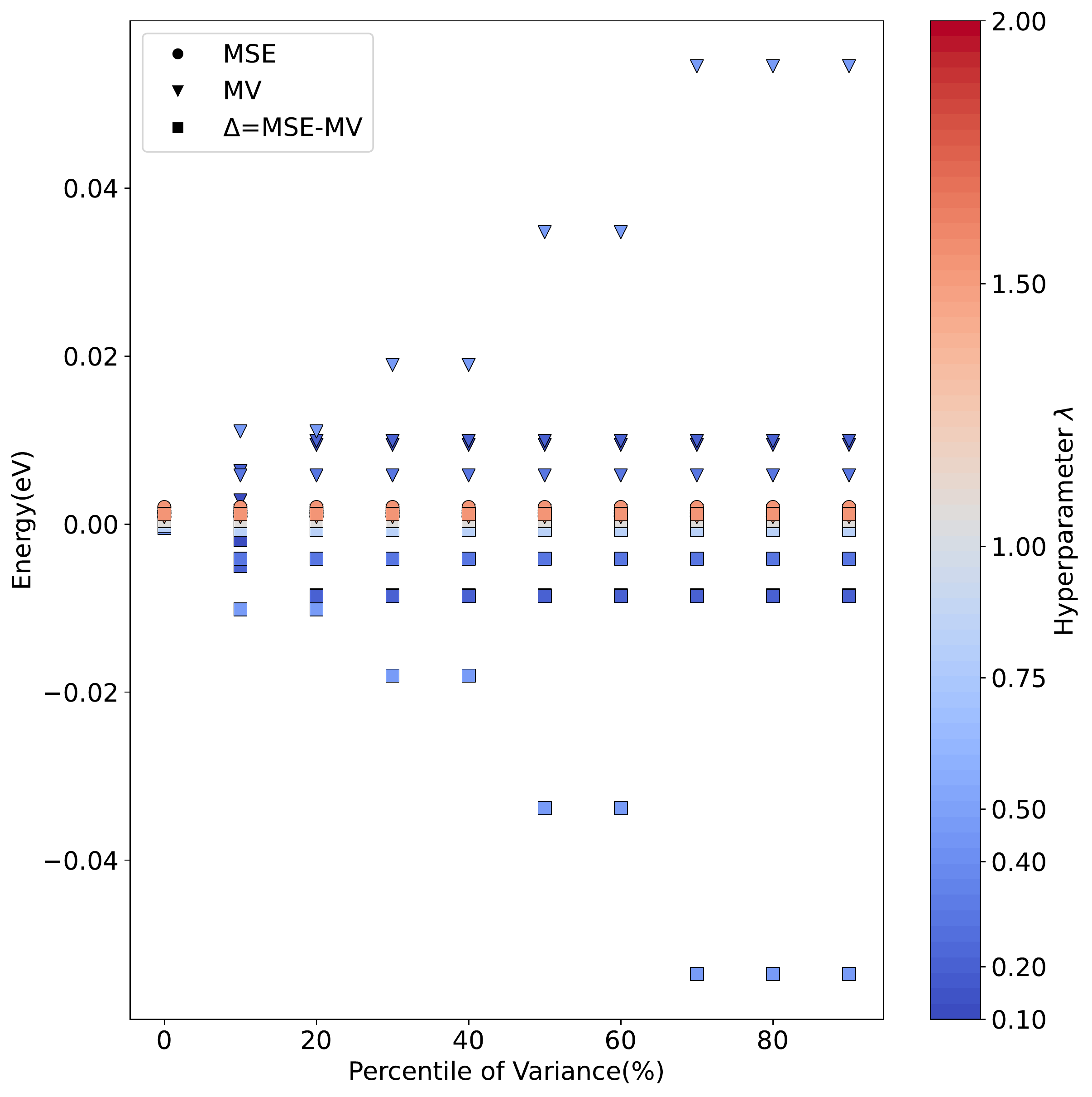}
    \caption{Difference between Mean Squared Error (MSE) and Mean
      Variance (MV) for different percentiles of the predicted
      variance. Values for MSE, MV and its difference ($\Delta$) at a
      given quantile are shown with different labels. The color bar
      indicates the values of the hyperparameter $\lambda$. The
      different dots are colored accordingly to its $\lambda$
      value. Value for 0.01 was excluded for clarity. }
\label{sifig:error_var_diff}
\end{figure}

\begin{figure}
    \centering
    \includegraphics[scale=0.7]{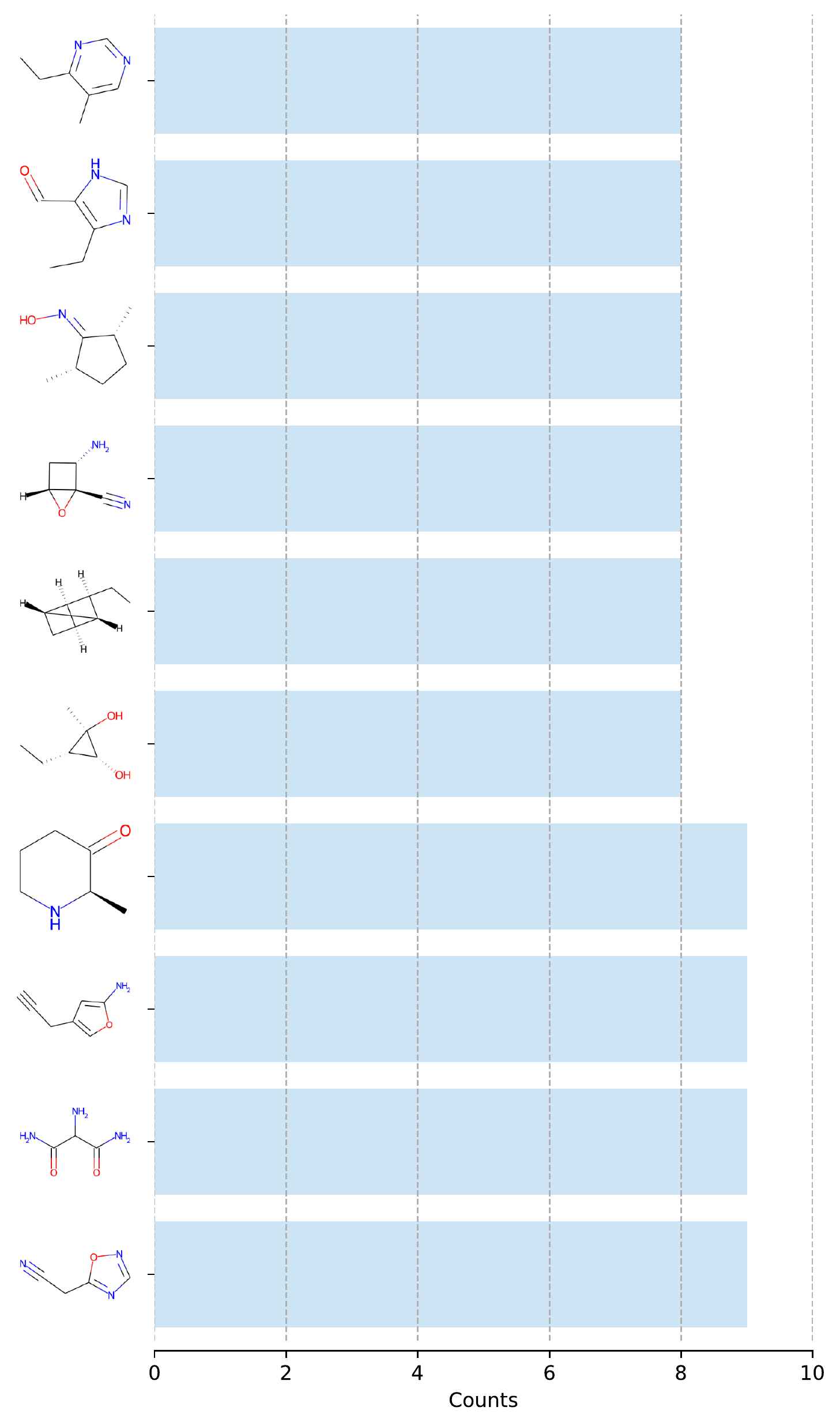}
    \caption{Top 10 Common True Positive (TP) molecules
      ($\varepsilon_{i}>\varepsilon_{max}$ and
      $\sigma_{i}>\sigma_{max}$). The $x-$axis shows how often a
      molecule appear as TP for the different values of hyperparameter
      $\lambda$, the $y-$axis show the characteristic chemical
      structure. There are only show molecules that appear for at
      least four different values of $\lambda$ }
\label{sifig:TP}
\end{figure}

\begin{figure}
    \centering
    \includegraphics[scale=0.7]{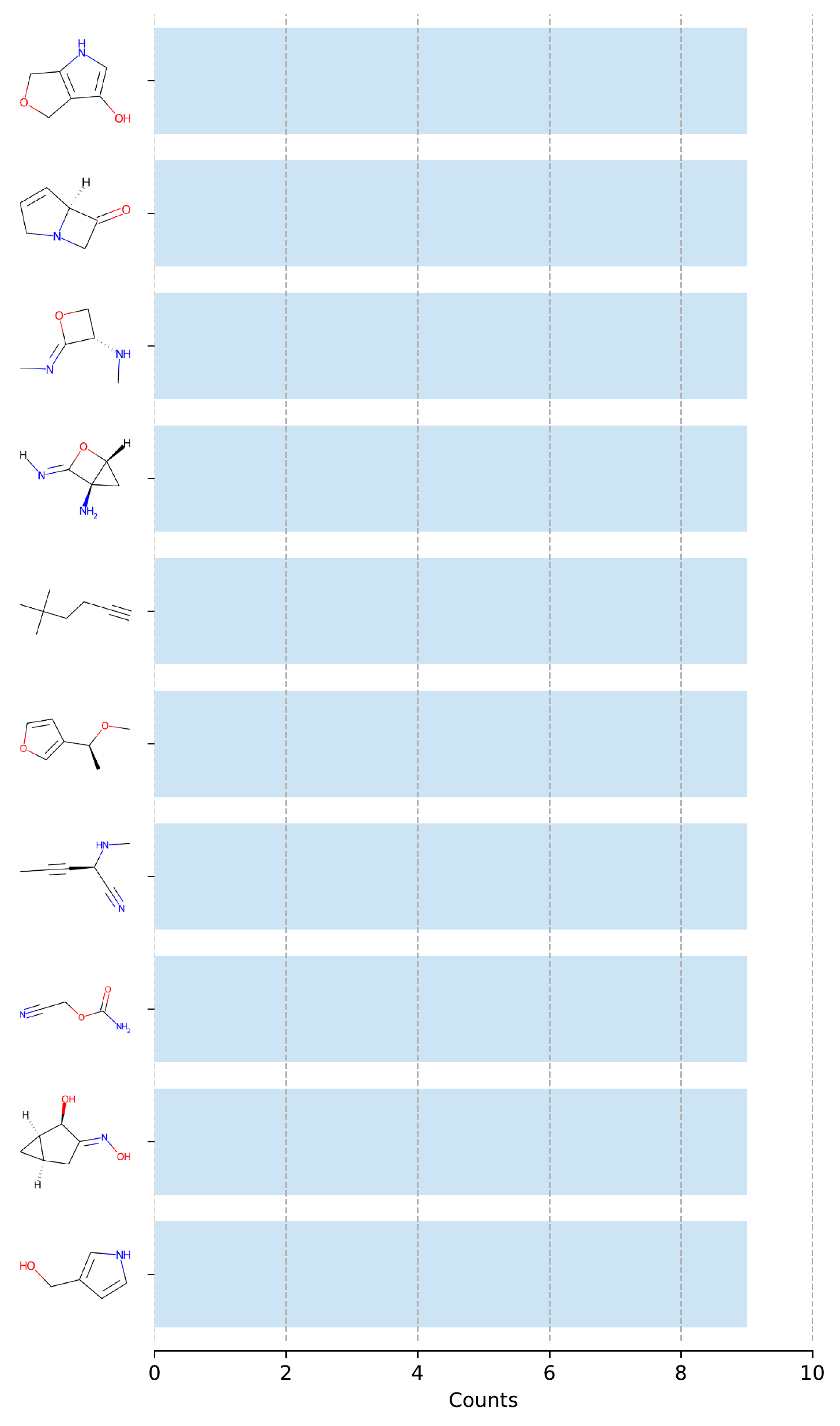}
    \caption{Top 10 Common True Negative (TN) molecules
      ($\varepsilon_{i}<\varepsilon_{max}$ and
      $\sigma_{i}<\sigma_{max}$). The $x-$axis shows how often a
      molecule appear as TN for the different values of hyperparameter
      $\lambda$, the $y-$axis show the characteristic chemical
      structure.}
\label{sifig:TN}
\end{figure}

\begin{figure}
    \centering
    \includegraphics[scale=0.7]{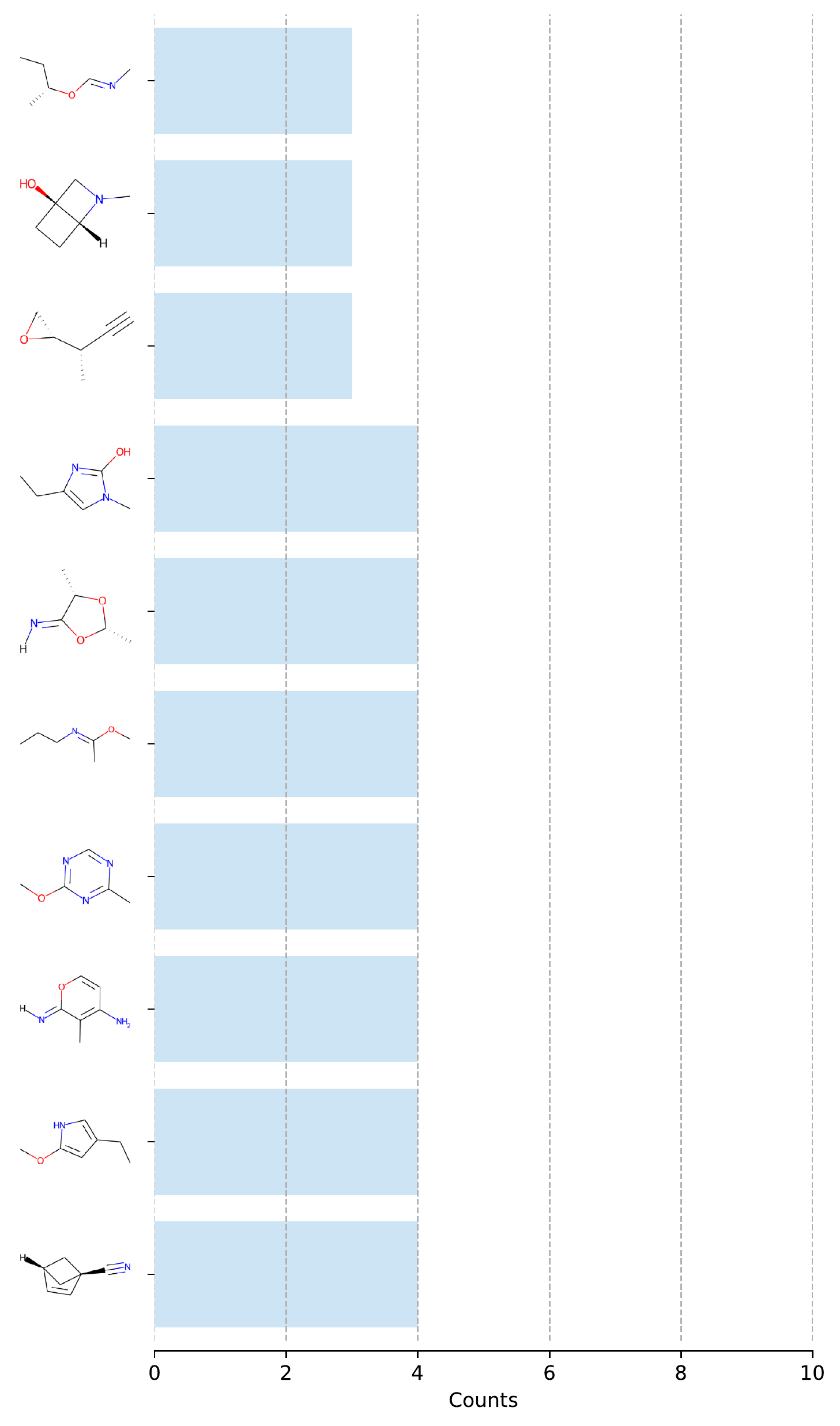}
    \caption{Top 10 Common False Positive (FP) molecules
      ($\varepsilon_{i}<\varepsilon_{max}$ and
      $\sigma_{i}>\sigma_{max}$). The $x-$axis shows how often a
      molecule appear as FP for the different values of hyperparameter
      $\lambda$, the $y-$axis show the characteristic chemical
      structure. }
\label{sifig:FP}
\end{figure}

\begin{figure}
    \centering
    \includegraphics[scale=0.7]{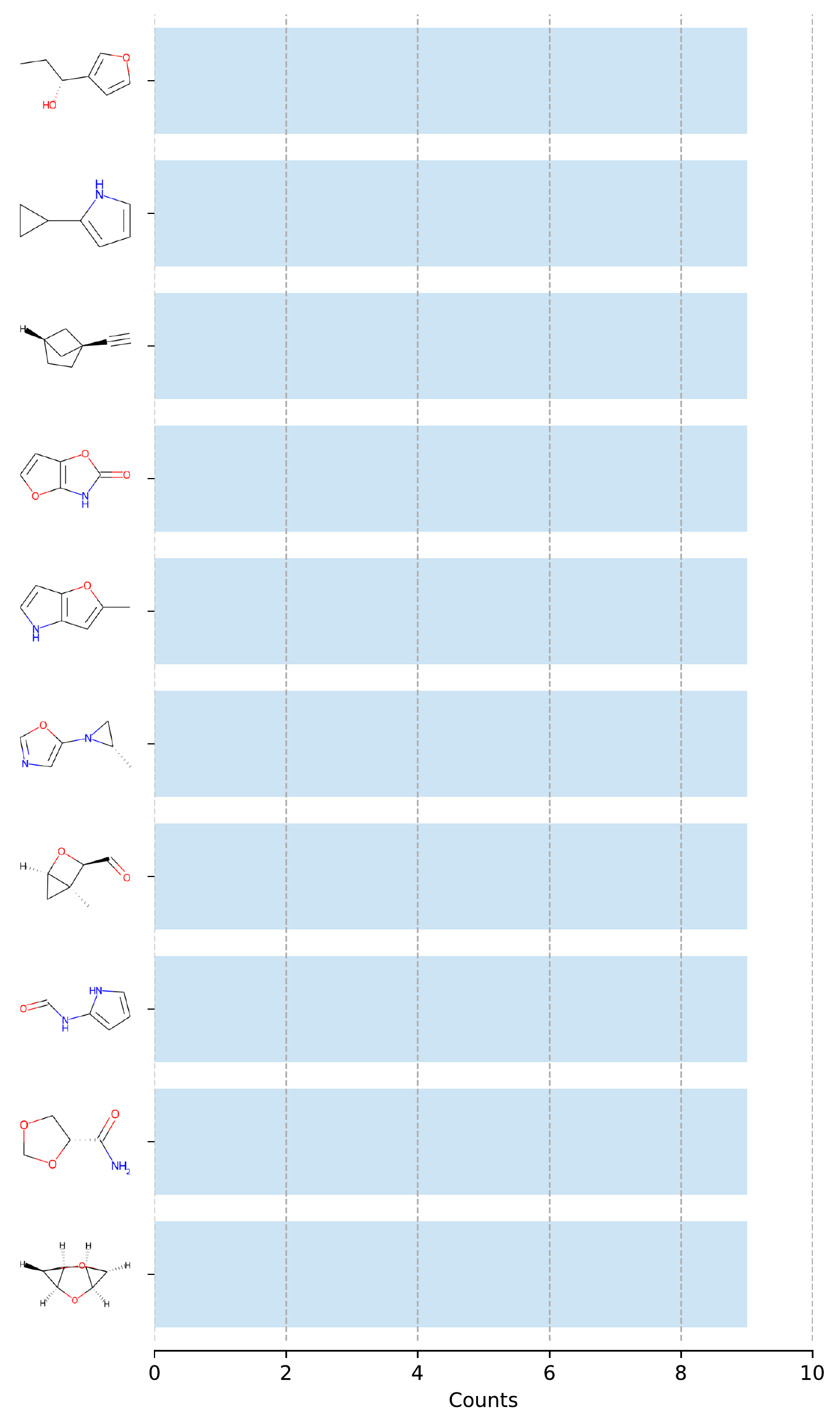}
    \caption{Top 10 Common False Negative (FN) molecules
      ($\varepsilon_{i}>\varepsilon_{max}$ and
      $\sigma_{i}<\sigma_{max}$). The $x-$axis shows how often a
      molecule appear as FN for the different values of hyperparameter
      $\lambda$, the $y-$axis show the characteristic chemical
      structure. }
\label{sifig:FN}
\end{figure}

\begin{figure}
    \centering
    \includegraphics[scale=0.45]{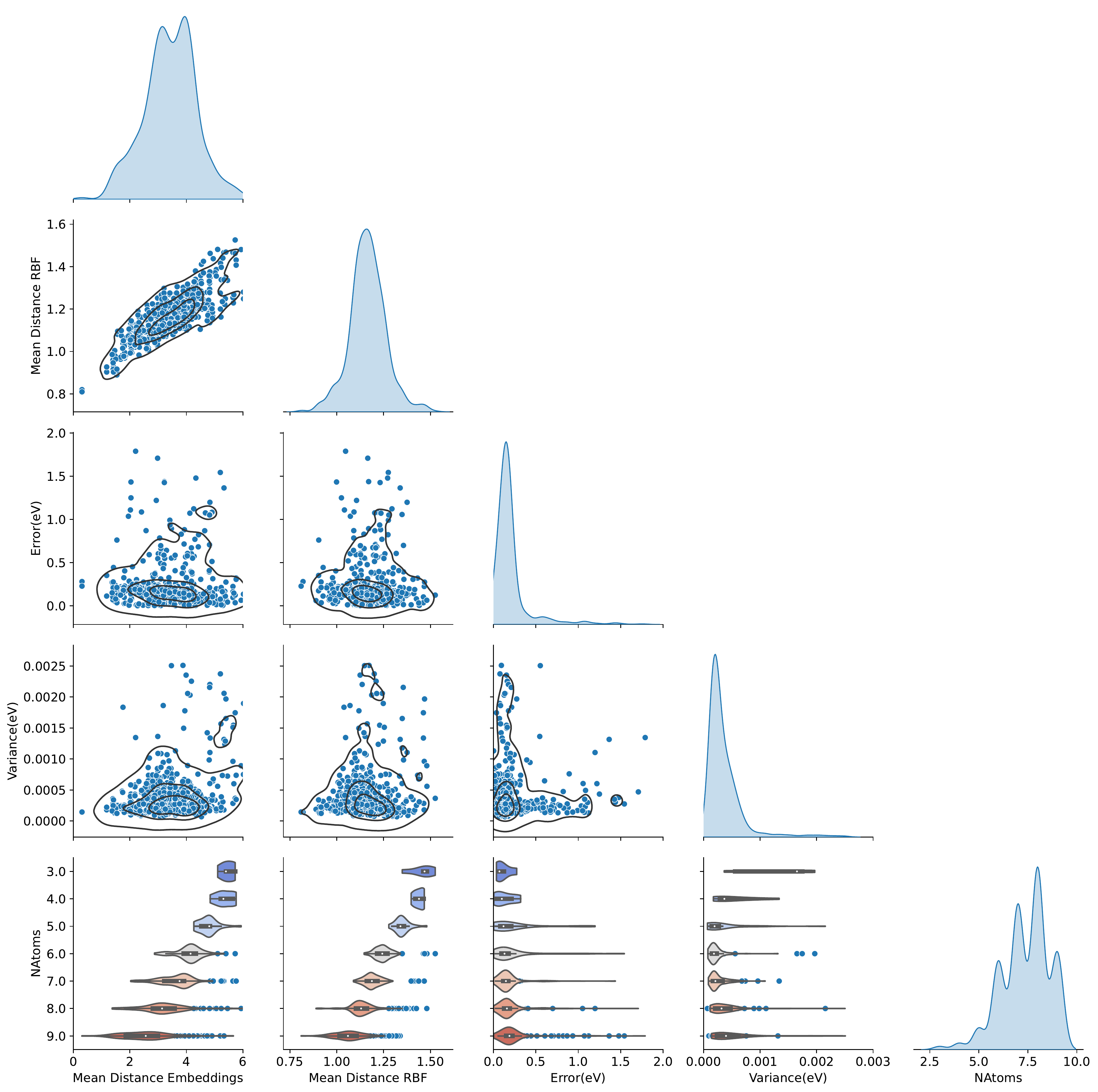}
    \caption{Overview of the results for the evaluation of molecules
      on the tautobase for $\lambda=0.2$. The diagonal of the figure
      shows the kernel density estimate of the considered properties
      (Mean Distance Embeddings, Mean Distance RBF, Error
      (eV),Variance (eV) and Number of Atoms). For each of the panels
      a correlation plot between the variable and a 2D kernel density
      estimate is shown. In the last row, violin plots for the
      different considered properties with respect to the number of
      atoms is shown.}
\label{sifig:pairplot02}
\end{figure}

\begin{figure}
    \centering
    \includegraphics[scale=0.45]{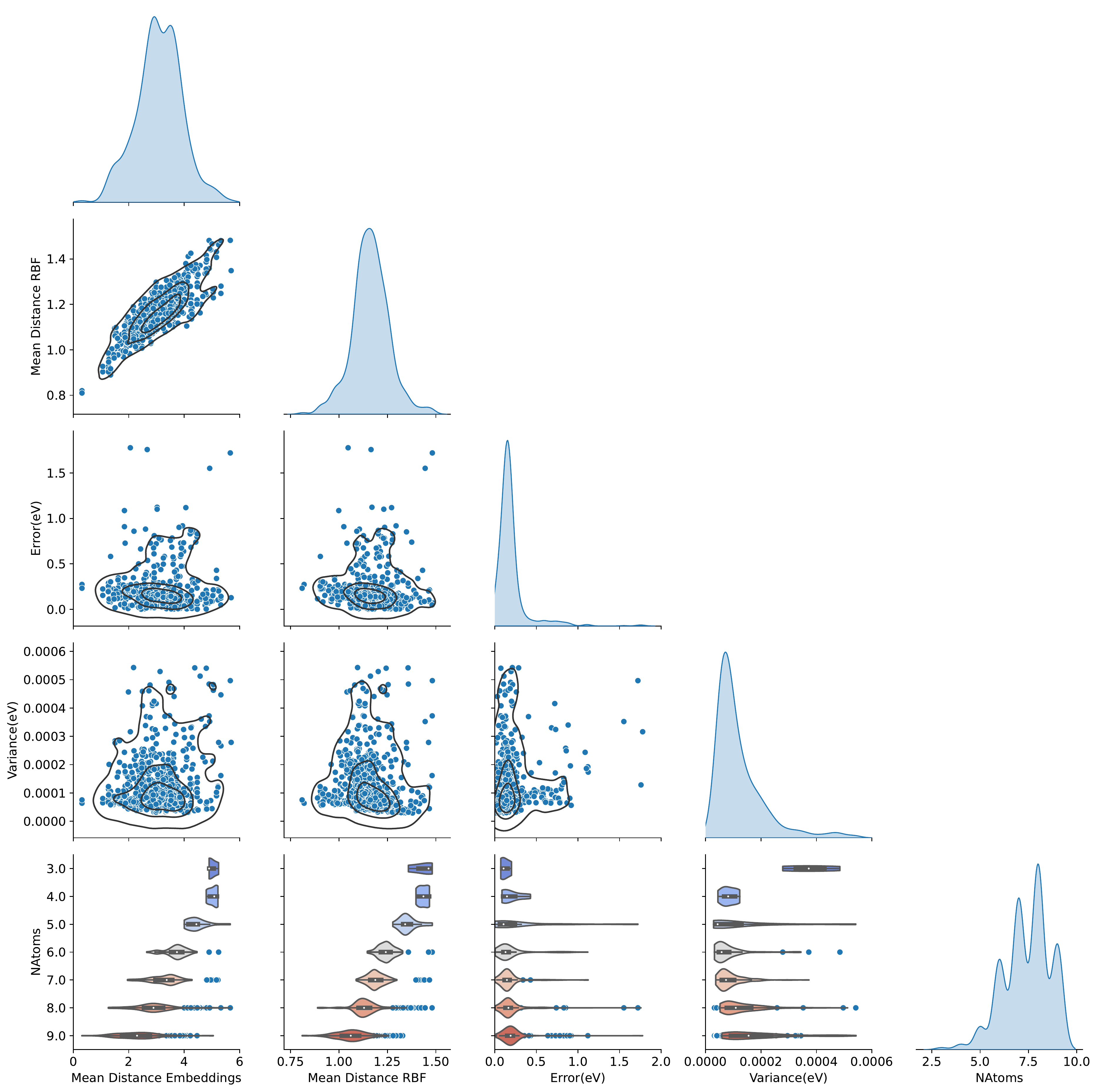}
    \caption{Overview of the results for the evaluation of molecules
      on the tautobase for $\lambda=0.4$. The diagonal of the figure
      shows the kernel density estimate of the considered properties
      (Mean Distance Embeddings, Mean Distance RBF, Error (eV),
      Variance (eV) and Number of Atoms). For each of the panels a
      correlation plot between the variable and a 2D kernel density
      estimate is shown. In the last row, violin plots for the
      different considered properties with respect to the number of
      atoms is shown.}
\label{sifig:pairplot04}
\end{figure}

\begin{figure}
    \centering
    \includegraphics[scale=1]{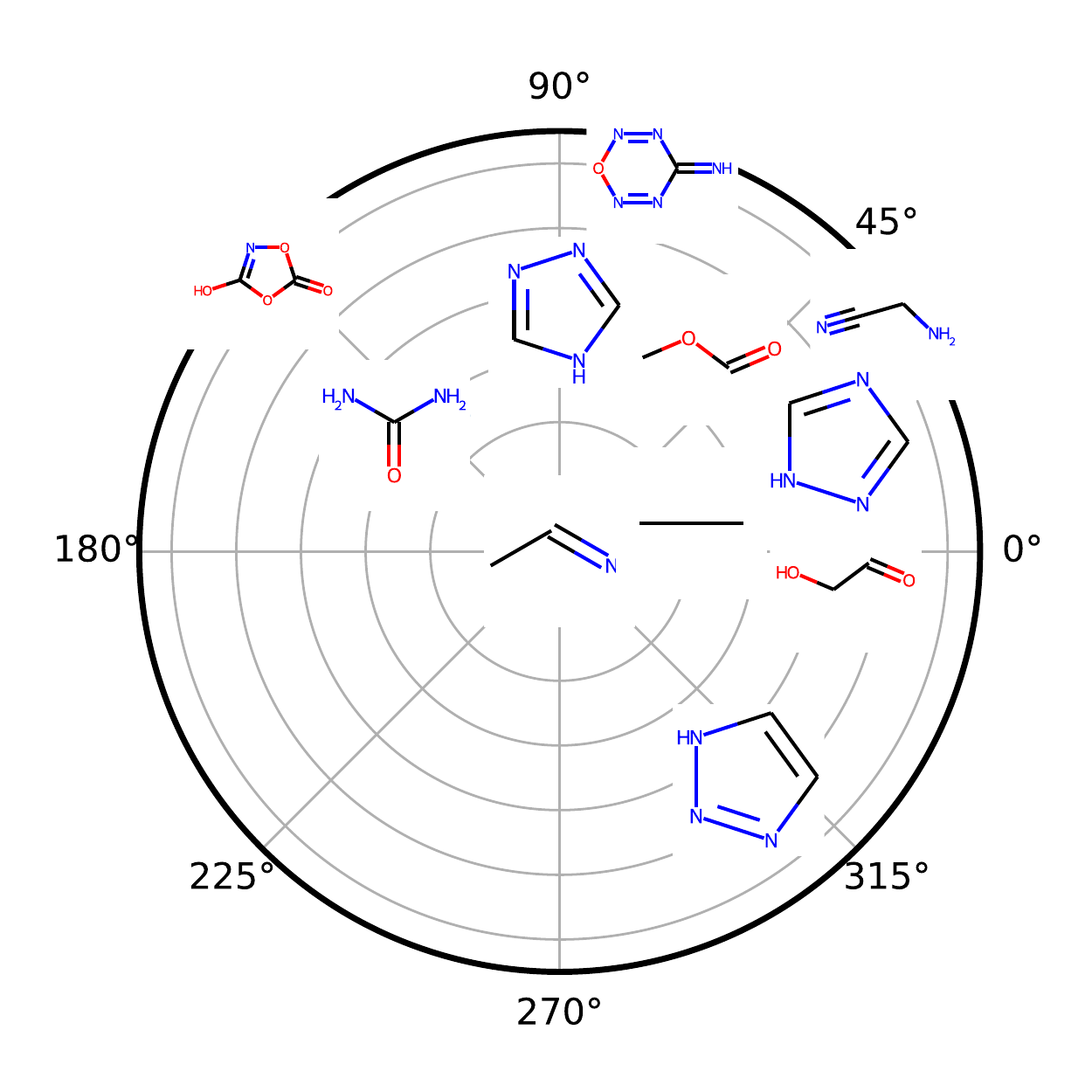}
    \caption{Polar plot of the mean distance between a molecule in the
      test set (Ethanimine) and molecules on the training set. The
      molecule in the center corresponds to the sample with the
      largest mean distance in RBF and embedding space.}
\label{sifig:polar_maxdist}
\end{figure}

\begin{figure}
    \centering
    \includegraphics[scale=1]{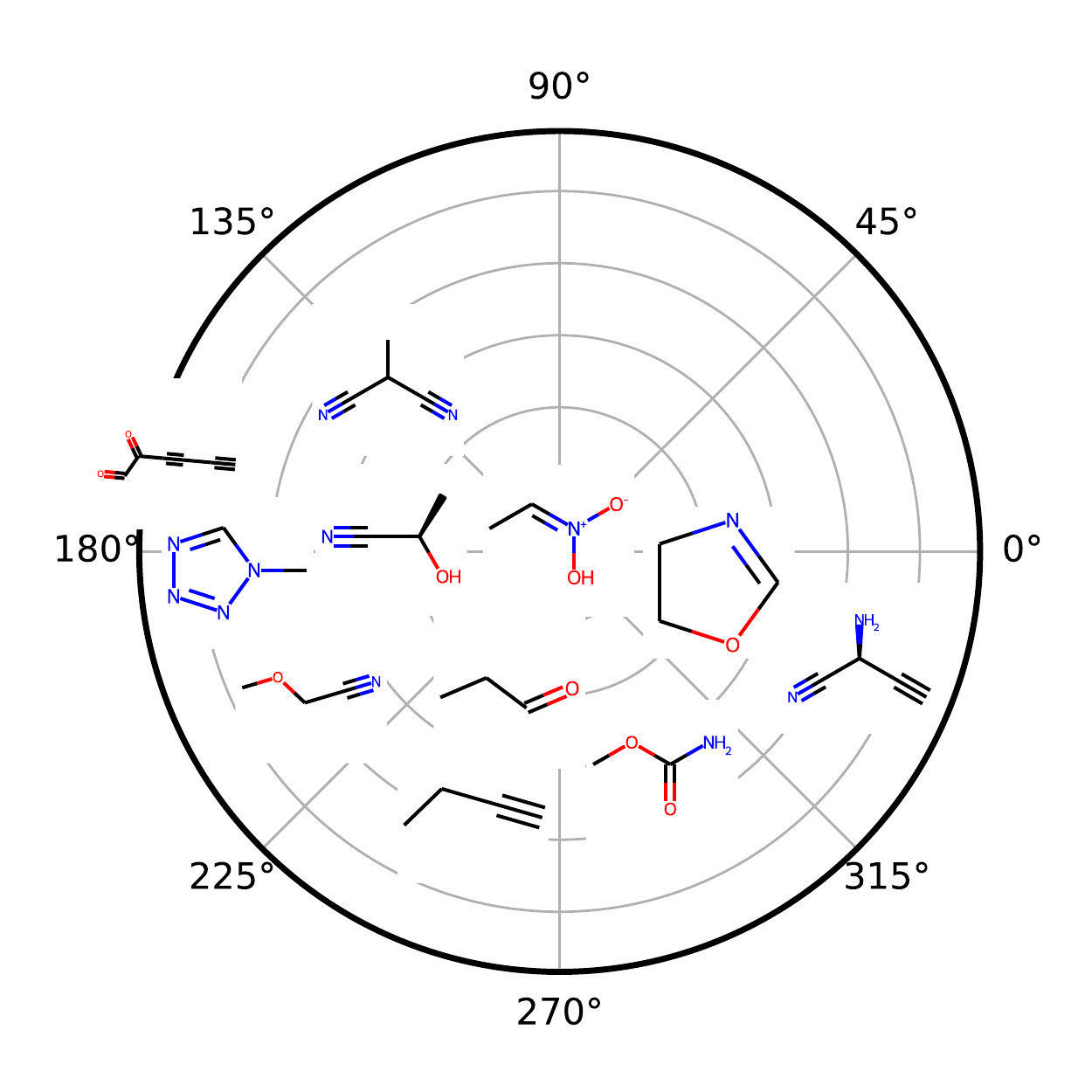}
    \caption{Polar plot of the mean distance between a molecule in the
      test set ((1E)-Ethylideneazinic acid) and molecules in the
      training set. The molecule in the center corresponds to the
      molecule in the test set with the largest prediction error at
      the 95th percentile.}
\label{sifig:Polar_maxerror}
\end{figure}

\begin{figure}
    \centering
    \includegraphics[scale=1]{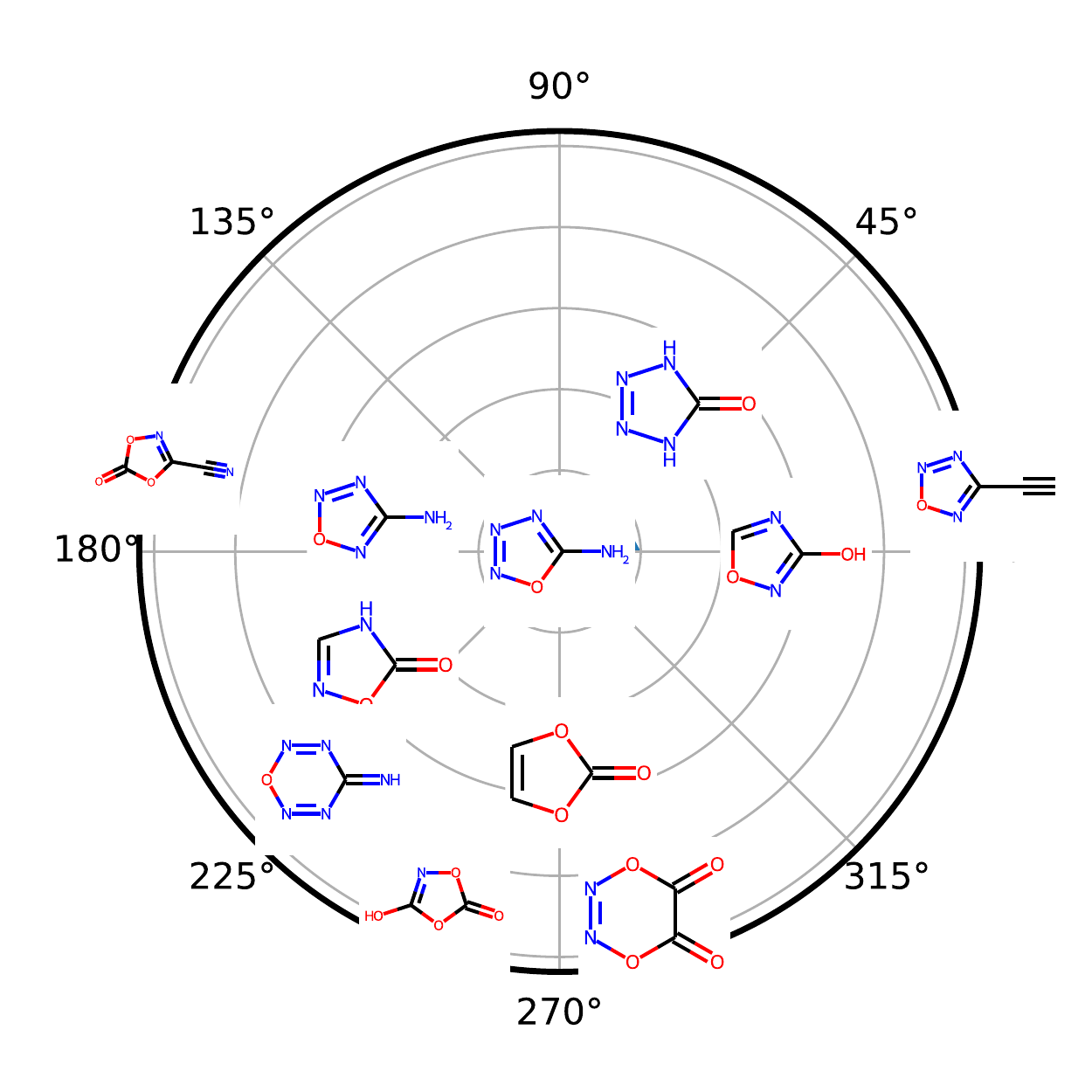}
    \caption{Polar plot of the mean distance between a molecule in the
      test set (1,2,3,4-Oxatriazol-5-amine) and molecules in the
      training set. The molecule in the center corresponds to the
      molecule in the test set with the largest predicted variance at
      95th percentile.}
\label{sifig:polar_maxvar}
\end{figure}

\begin{figure}
    \centering \includegraphics{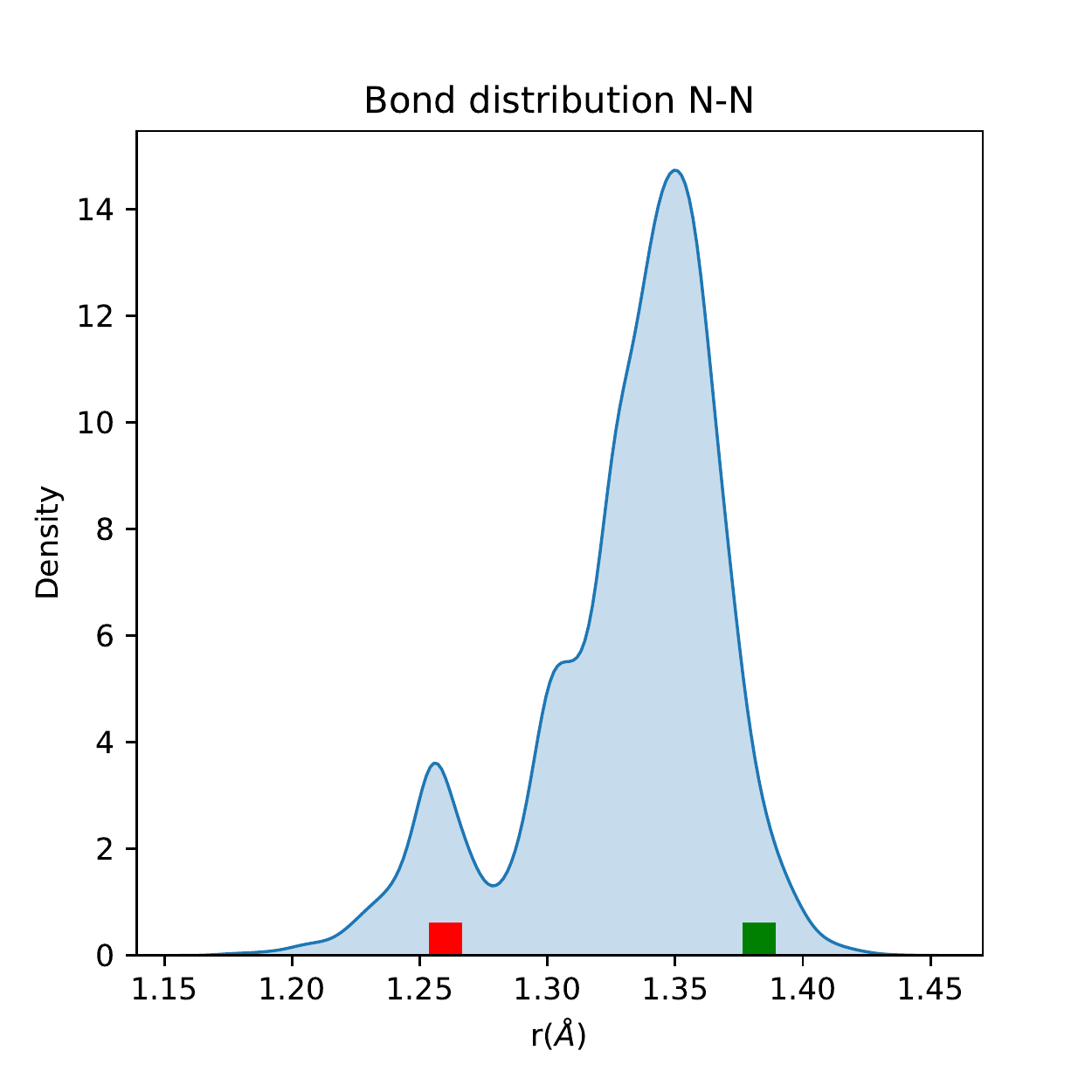}
    \caption{N-N bond distribution from all molecules in the QM9
      database\cite{ramakrishnan2014quantum}. The red point indicates
      the N-N bond distance for molecule A1 in Figure
      \ref{fig:tautomers}B and the green point indicates that for
      molecule B1.}
\label{sifig:bond_dist}
\end{figure}

\clearpage

\bibliography{aipsamp}

\end{document}